\RequirePackage{fix-cm}
\documentclass[twocolumn,epjc3]{svjour3}  

\makeatletter
\expandafter\let\csname opt@amsmath.sty\endcsname\relax
\makeatother

\RequirePackage{graphicx}
\RequirePackage[numbers,sort&compress]{natbib}
\RequirePackage[colorlinks,citecolor=blue,urlcolor=blue,linkcolor=blue]{hyperref}

\RequirePackage{amsmath, amssymb, dsfont, physics}
\RequirePackage{xcolor}
\RequirePackage{ulem} % for sout
\graphicspath{{figures/}}

\definecolor{darkgreen}{HTML}{006400}

\newcommand{\pdagger}{{\phantom{\dagger}}}

\journalname{Eur. Phys. J. B}

\begin{document}

\title{Moments and multiplets in moir\'e materials} %\thanksref{t1}}
\subtitle{A pseudo-fermion functional renormalization group for spin-valley models}


\author{Lasse Gresista\thanksref{e1}
        \and
        Dominik Kiese %\thanksref{addr1} %etc.
        \and
        Simon Trebst %\thanksref{addr1}
}

%\thankstext[$\star$]{t1}{Thanks to the title}
\thankstext{e1}{e-mail: gresista@thp.uni-koeln.de}


\institute{Institute for Theoretical Physics, University of Cologne, 50937 Cologne, Germany}

\date{Received: date / Accepted: date}

\abstractdc{
The observation of strongly-correlated states in moir\'e systems has renewed the conceptual interest 
in magnetic systems with higher SU(4) spin symmetry, e.g.\ to describe Mott insulators where the 
local moments are coupled spin-valley degrees of freedom.
Here, we discuss a numerical renormalization group scheme to explore the formation of spin-valley ordered  and unconventional spin-valley liquid states at zero temperature based on a pseudo-fermion representation. 
Our generalization of the conventional pseudo-fermion functional renormalization group approach for $\mathfrak{su}$(2) spins is
capable of treating diagonal and off-diagonal couplings of generic spin-valley exchange Hamiltonians in the self-conjugate representation of the $\mathfrak{su}$(4)
algebra. To achieve proper numerical efficiency, we derive a number of symmetry constraints on the flow equations
that significantly limit the number of ordinary differential equations to be solved.
As an example system, we investigate a diagonal SU(2)$_{\textrm{spin}}$ $\otimes$ U(1)$_{\textrm{valley}}$ model on the triangular lattice which exhibits a rich phase diagram of spin and valley ordered phases.}

\maketitle

%%%%%%%%%%%%%%%%%%%%%%%%%%%%%%%%%%%%%%%%%%%%%%%%%%%%%%%%%%

\section{Introduction}
\label{sec:introduction}

Moir\'e materials that exhibit flat bands such as twisted bilayer graphene (tBG) or certain van der Waals heterostructures such as hexagonal boron nitride (TLG/h-BN) have recently been established as  novel, highly tunable platforms for the study of strongly correlated electrons. Relative to an almost vanishing bandwidth, residual interactions in these materials can induce a plethora of different many-body phenomena ranging from the formation of correlated insulators \cite{cao2018, chen2020, liu2020, cao2020} and superconductors \cite{cao2018a, lu2019, chen2019} to anomalous quantum Hall effects \cite{sharpe2019}. However, a microsopic description of these phenomena is a formidable challenge as the number of of low-energy degrees of freedom is often increased
\cite{koshino2018,xian2019, zhang2019} in comparison to conventional Mott insulators.

More specifically, it has been argued \cite{xu2018, yuan2018}, that multi-orbital Hubbard models can describe the flat band physics in e.g. TLG/h-BN within the topologically trivial regime, where fully symmetric Wannier states may be constructed \cite{po2018}. The proposed interaction terms for the corresponding Hamiltonians usually include a generalized Hubbard $U$ \cite{xu2018, yuan2018, classen2019} as well as Hund's type couplings.
Performing a strong coupling expansion where one treats the interactions as the dominant energy scale, these extended Hubbard models can then be mapped to $\mathfrak{su}$(4)\footnote{With $\mathfrak{su}$(4) we refer to the Lie algebra of the Lie group SU(4).} spin-valley Hamiltonians that may be used as a starting point to investigate the nature of the correlated insulating states. 
The so-derived $\mathfrak{su}$(4) models bear a close resemblance to  Kugel-Khomskii models \cite{kugel1982}
that have a long history in the study of transition metal oxides, where they are used to capture the Jahn-Teller physics of intertwined spin and 
orbital degrees of freedom. 
Increasing the number of relevant microscopic degrees of freedom (in comparison to conventional quantum magnets) has been particlularly appreciated to boost quantum fluctuations independent of, e.g., lattice geometries \cite{feiner1997}, which has made Kugel-Khomskii models a recurring target in the search for unusual many-body states such as quantum spin-orbital liquids \cite{corboz2012, natori2018, natori2019, kiese2020a}.
As such, one might expect the $\mathfrak{su}$(4) spin-valley physics relevant to the correlated insulating states of moir\'e materials to hold similar promise for the observation of spin-valley liquid states with macroscopic entanglement and potentially long-range, topological order.

In this manuscript, we present a powerful numerical scheme to analyze such $\mathfrak{su}$(4) spin-valley (or spin-orbital) models 
based on a functional renormalization group (FRG) technique. 
Our approach is based on the pseudo-fermion FRG (pf-FRG) \cite{reuther2010}, approximating the elementary spin operators of the 
six-dimensional, self-conjugate representation of $\mathfrak{su}$(4) by auxiliary complex fermions combined with an on-average constraint on the number of particles per site. Our approach allows to go beyond mean-field level by treating competing instabilities in different interaction channels on equal footing, and is able to capture both, long-ranged spin and/or valley ordered states as well as spin-valley liquid phases. In expanding previous work (by some of us) \cite{kiese2020a}, we extend the range of applicability of this approach to models with off-diagonal interactions in either spin or valley space by formulating an efficient vertex parametrization derived from a meticulous symmetry analysis. We demonstrate the feasibility of this method by studying a spin-valley Hamiltonian with SU(2)$_{\textrm{spin}}$ $\otimes$ U(1)$_{\textrm{valley}}$ symmetry where we identify a plethora of spin and valley orderded phases from a state-of-the-art numerical implementation of pf-FRG \cite{thoenniss2020, kiese2021}.

The remainder of this manuscript is structured as follows. To begin with, we introduce the spin-valley Hamiltonian of interest on a general level and discuss its specific form for TLG/h-BN as a concrete example in Sec.~\ref{sec:model}. We will continue by reviewing the pf-FRG approach (Sec.~\ref{sec:overview}), its generalization for $\mathfrak{su}$(4) models as well as the implementation of model specific symmetries (Sec.~\ref{sec:symmetry-classification}). Finally, numerical results for the phase diagram of a SU(2)$_{\textrm{spin}}$ $\otimes$ U(1)$_{\textrm{valley}}$ model on the triangular lattice are presented and examined in Sec.~\ref{sec:results}.

\section{Model}
\label{sec:model}

\begin{figure*}
    \centering
    \includegraphics{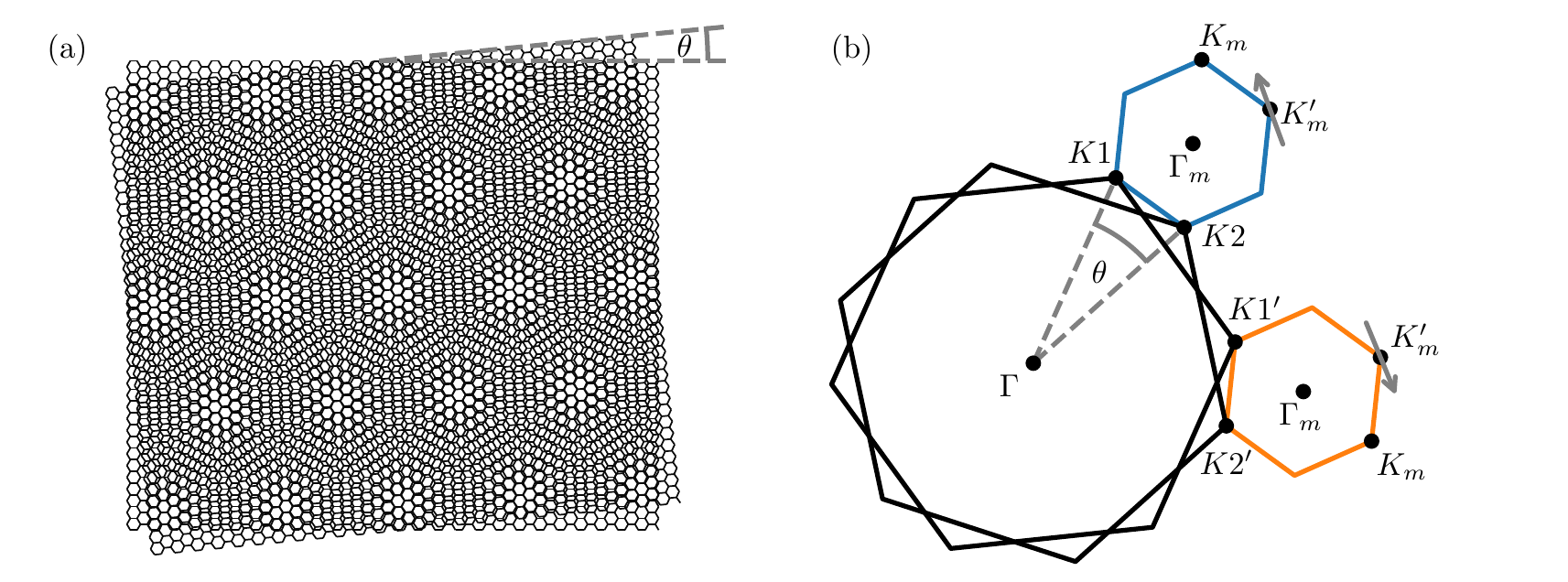}
    \caption{(a) Moir\'e pattern emerging in two stacked layers of graphene with a relative twist angle $\theta$. Clearly visible are the different regions with AA, BA and BB stacking leading to a triangular super-lattice structure.  (b) Construction of the two degenerate mini Brillouin zones from the difference of the K (or K') points of the two layers of graphene. Additional to the spin degree of freedom, indicated by the grey arrows, the electrons obtain a valley degree of freedom due to the possibility of being in either one of the mini Brillouin zones at the two valleys (at the K and K' points) of the graphene band structure.}
    \label{fig:tbg-visualization}
\end{figure*}

Microscopically, the SU(4) models of interest in this manuscript can be cast in terms of a general Hamiltonian
\begin{equation}
    \label{eq:su4-hamiltonian}
    \mathcal{H} = \frac{1}{8} \sum_{\langle ij \rangle} J (1 + \boldsymbol{\sigma}_i \boldsymbol{\sigma}_j) (1 + \boldsymbol{\tau}_i \boldsymbol{\tau}_j) \,,
\end{equation}
that couples two elementary $\mathfrak{su}$(2) degrees of freedom, captured by the operators $\sigma$ and $\tau$, which might denote a spin and valley (or oribtal) degree of freedom. The overall SU(4) symmetry of the Hamiltonian arises from the balanced couplings of equal strength in both degrees of freedom, i.e. $J$ is identical for the Heisenberg-like coupling of spins $\boldsymbol{\sigma}_i \boldsymbol{\sigma}_j$ on sites $i$ and $j$ (with $\boldsymbol{\sigma}_i = (\sigma_j^x, \sigma_j^y, \sigma_j^z)^T$) and a similar interaction of the valley degrees of freedom $\boldsymbol{\tau}_i \boldsymbol{\tau}_j$. 
Such valley degrees of freedom arise, in the context of tBG and related moir\'e materials, from the Dirac cones in the original graphene bands, which hybridize between the two layers upon twisting and thereby add an extra index \cite{bistritzer2011} to the moir\'e bands, as illustrated in Fig.~\ref{fig:tbg-visualization}. Before drawing broad attention in the context of moir\'e materials, the spin-orbital variant of this model has been widely studied as Kugel-Khomskii model \cite{kugel1982}, often in connection with Jahn-Teller physics in transition metal oxides where spin and orbital ordering are intertwined \cite{Khomskii2014}.
We note that while we will frame our discussion of the SU(4) model \eqref{eq:su4-hamiltonian} in the language of spin-valley physics relevant to moir\'e materials, the presented pf-FRG approach is equally applicable in the study of such spin-orbital models. We will return to this point in the discussion section at the end.

In what we will discuss in the following, we will put a focus on the self-conjugate representation of $\mathfrak{su}$(4), where the spin-valley operators can be represented in terms of fermionic creation and annihilation operators as
\begin{align}
    \sigma_{i}^{\mu} \tau_{i}^{\kappa} &\equiv \sigma_{i}^{\mu} \otimes \tau_{i}^{\kappa} = f_{i s l}^{\dagger} \theta_{s s'}^{\mu} \theta^{\kappa}_{l l'} f^{\pdagger}_{i s' l'} \notag \\ 
    \sigma_i^{\mu} &\equiv \sigma_i^{\mu} \otimes \tau_i^0 = f_{i s l}^{\dagger} \theta_{s s'}^{\mu} f^{\pdagger}_{i s' l} \notag \\
    \tau_i^\kappa &\equiv \sigma_i^0 \otimes \tau_i^{\kappa} = f_{i s l}^{\dagger} \theta_{l l'}^{\kappa} f^{\pdagger}_{i s l'} \,,
    \label{eq:spin-valley-operators}
\end{align}
with a local half-fillling constraint 
\begin{equation}
  f_{i sl}^{\dagger} f_{i sl}^{\pdagger} = 2
  \label{eq:half-filling}
\end{equation}
subject to every lattice site, where summation over repeated spin indices $s$ and valley indices $l$ is implied. Here, $\theta^{\mu}$ denotes a Pauli matrix with $\mu \in \{0, 1, 2, 3\}$ and $\theta^{0} = \mathds{1}$. Allowing also for more generic, i.e. SU(4) breaking, interactions, any bilinear spin-valley Hamiltonian can be written as 
\begin{equation}
    \label{eq:hamiltonian}
    \begin{aligned}
    \mathcal{H} &= \frac{1}{8}\sum_{ij}\left[ (\sigma_i^\mu J_{s,ij}^{\mu\nu} \sigma_j^\nu) (\tau_i^\kappa J_{v,ij}^{\kappa\lambda} \tau_j^\lambda) + I_{ij} \hat{n}_i \hat{n}_j\right]\\
    & \equiv \frac{1}{8} \sum_{ij} \left[
    (\sigma_i^\mu \otimes \tau_i^\kappa)
    \left( 
    J_{s, ij}^{\mu \nu} \otimes
    J_{v, ij}^{\kappa \lambda}
    \right)
    (\sigma_j^\nu \otimes \tau_j^\lambda) + I_{ij} \hat{n}_i \hat{n}_j\right] \,,
    \end{aligned} 
\end{equation}
where $J_{s, ij}^{\mu \nu} \otimes J_{v, ij}^{\kappa \lambda}$ is understood as the Kronecker product of the spin and valley exchange matrices and summation over repeating $\mu,\nu,\kappa$ or $\lambda$ is again implied. Here, $\hat{n}_i$ is the density operator $\hat{n}_i \equiv \sigma_i^0 \tau_i^0 = f^\dagger_{isl}f^{\pdagger}_{isl}$, and the term proportional to the coupling $I_{ij}$ is needed to potentially cancel the density term $\sim \sigma_i^0 \tau_i^0  J^{00}_{s, ij}J^{00}_{v, ij}  \sigma_j^0  \tau_j^0$, which does not appear in pure $\mathfrak{su}(4)$ spin models as, e.g., the SU(4) symmetric Hamiltonian in Eq.~\eqref{eq:su4-hamiltonian}.

To keep the numerical effort for employing our pf-FRG approach at a manageable level, we assume a specific form of the exchange matrices, namely, that both, the spin and the valley exchange only couple bilinears of spin/valley or density operators and that the spin exchange is $\mathds{Z}_2 \times \mathds{Z}_2 \times \mathds{Z}_2$ symmetric, thus
\begin{equation}
    \begin{aligned}
        \label{eq:coupling-matrices}
        J_{s, ij} &= \begin{pmatrix}
            J_{s,ij}^d & 0            & 0            & 0 \\
            0        & J_{s,ij}^{x} & 0            & 0 \\
            0        & 0            & J_{s,ij}^{y} & 0 \\
            0        & 0            & 0            & J_{s,ij}^{z}
        \end{pmatrix}\\
        J_{v, ij} &= \begin{pmatrix}
            J_{v,ij}^d & 0 & 0 & 0 \\
            0 & J_{v,ij}^{xx} & J_{v,ij}^{xy} & J_{v,ij}^{xz} \\
            0 & J_{v,ij}^{yx} & J_{v,ij}^{yy} & J_{v,ij}^{yz} \\
            0 & J_{v,ij}^{zx} & J_{v,ij}^{zy} & J_{v,ij}^{zz}
        \end{pmatrix} \,.
    \end{aligned}
\end{equation}
This form, although it spoils the generality of Eq.~\eqref{eq:hamiltonian}, is nevertheless relevant to certain practical applications. For instance, the effective Hamiltonian for TLG/h-BN  \cite{zhang2019} can be recast to this form. Originally, the former is often given as 
\begin{equation}
    \begin{aligned}
        \mathcal{H} &= \frac{J_1}{8} \sum_{\langle ij \rangle} (1 + \boldsymbol{\sigma}_i \boldsymbol{\sigma}_j)(1 + \boldsymbol{\tau}_i \boldsymbol{\tau}_j) \\
    &+ \frac{J_2}{8} \sum_{\langle\langle ij \rangle\rangle} (1 + \boldsymbol{\sigma}_i \boldsymbol{\sigma}_j)(1 + \boldsymbol{\tau}_i \boldsymbol{\tau}_j) \\
    &+ \frac{1}{8} \sum_{\langle ij \rangle} J^1_{p;ij}(1 + \boldsymbol{\sigma}_i \boldsymbol{\sigma}_j)(\tau^x_i \tau^x_j +\tau^y_i \tau^y_j)\\
    &+ \frac{1}{8} \sum_{\langle ij \rangle} J^2_{p;ij}(1 + \boldsymbol{\sigma}_i \boldsymbol{\sigma}_j)(\tau^x_i \tau^y_j - \tau^y_i \tau^x_j)\\
    &+ O\left(\frac{t^3}{U^2}\right),
    \end{aligned}
\end{equation}
which, in addition to SU(4) symmetric nearest neighbour ($\sim J_1$) and next-nearest neighbour ($\sim J_2$) interactions, contains both diagonal $\sim J^1_{p, ij}$ and off-diagonal $\sim J^2_{p,ij}$ valley exchange that breaks the SU(4) symmetry down to an SU(2)$_\mathrm{spin}$ $\otimes$ U(1)$_\mathrm{valley}$ symmetry. Comparing this model to the form of the general spin-valley Hamiltonian defined in Eq.~\eqref{eq:hamiltonian}, the nearest neighbour exchange matrices can be written as
\begin{equation}
    \begin{aligned}
        J_{s, ij} &= \mathds{1} \\
        J_{v, ij} &= \begin{pmatrix}
                J_1 & 0 & 0 & 0 \\
                0 & J_1 & J^2_{p;ij} & 0 \\
                0 & -J^2_{p;ij} & J_1 +  J^1_{p;ij} & 0 \\
                0 & 0 & 0 & J_1 +  J^1_{p;ij}
            \end{pmatrix}
            \, ,
    \end{aligned}
\end{equation}
and the next-nearest neighbour exchange is fully SU(4) symmetric, showing that they are indeed captured by the exchange matrices defined in Eq.~\eqref{eq:coupling-matrices}.

%%%%%%%%%%%%%%%%%%%%%%%%%%%%%%%%%%%%%%%%%%%%%%%%%%%%%%%%%%

\section{pf-FRG for spin-valley models: An overview}
\label{sec:overview}
We now proceed to the core methodological advancement of this manuscript, which will be laid out in this section -- the extension of the conventional pf-FRG to spin-valley models described by Hamiltonians of the form given in Eq.~\eqref{eq:hamiltonian}, with general, diagonal and off-diagonal couplings as defined by Eq~\eqref{eq:coupling-matrices}. 
To set the stage, we will first revisit the flow equations of the conventional pf-FRG approach for $\mathfrak{su}$(2) spins and explain how the numerical solution of the flow equations can be used to determine whether and what type of magnetic order forms for a particular spin Hamiltonian at zero temperatures. 
We then proceed to the adapted pf-FRG approach for spin-valley models, for which we derive an efficient parametrization of the self-energy and two-particle vertex in what is a direct extension of the parametrization for $\mathfrak{su}(2)$ spin models with generic two-spin interactions \cite{buessen2019}. Our particular focus is on constraints that symmetries of the spin-valley Hamiltonian pose on the parametrized vertex functions  -- very similar to the $\mathfrak{su}(2)$ case but with slight differences which we especially highlight. To put these equations into numerical practice, we discuss our implementation of the spin-valley pf-FRG approach and its algorithmic scaling.
% and discuss the degree in which the half-filling constraint in Eq.~\eqref{eq:half-filling} is fulfilled. 
This section is intended as an overview stating the main results of our study important for the implementation of the pf-FRG for spin-valley models. Readers looking for a more detailed discussion of how the symmetries of the Hamiltonian lead to the parametrization and symmetry constraints are referred to Sec.~\ref{sec:symmetry-classification}.

\subsection{Pseudo fermion functional renormalization group}

Let us set the stage by revisiting some of the conceptual steps of the pseudo-fermion FRG, which has originally been formulated for bilinear $\mathfrak{su}(2)$ spin models \cite{reuther2010} with generic (diagonal and off-diagonal) interactions \cite{buessen2019} and later generalized to SU(N) Heisenberg models \cite{buessen2018a}, in the context of the spin-valley models at hand. By going to a pseudofermion representation of the original degrees of freedom, one arrives at a fermionic representation of the original model (with an additional half-filling constraint) as outlined in the previous section. One can then proceed to apply the well established methods of the fermionic FRG \cite{Wetterich1993,kopietz2010}.

An important distinction to electronic systems is that the pseudofermion Hamiltonian exhibits only a quartic interaction term and no quadratic kinetic terms. This readily implies that 
the free propagator is diagonal in all its arguments and takes the simple form
\begin{equation}
    G_0(1', 1) = G_0(\omega_1) \delta_{i_{1'} i_1} \delta_{s_{1'} s_1} \delta_{l_{1'}l_1} \delta_{\omega_{1'} \omega_1},
\end{equation} 
with $G_0(\omega) = (i\omega)^{-1}$. The multi-index $1 = (i_1, s_1, l_1, \omega_1)$ consists of a lattice site index $i_1$, a spin index $s_1$, a fermionic Matsubara frequency $\omega_1$ and, for spin-valley models, the additional valley index $l_1$. To implement the RG scale, or cutoff, $\Lambda$ we multiply a regulator to the free propagator
\begin{equation}
    \label{eq:cutoff}
	G_0^\Lambda(\omega) = G_0(\omega) (1-e^{-\omega^2/\Lambda^2}),
\end{equation}
where we choose a smooth regulator for improved numerical stability. The pf-FRG flow equations are then given as a special case of the general fermionic FRG equations by assuming that the flowing self-energy is, just as the free propagator, diagonal in all its arguments. This assumption is true for arbitrary spin-models bilinear in $\mathfrak{su}(2)$ spin operators \cite{buessen2019}. For spin-valley Hamiltonians, however, we will show in Sec.~\ref{sec:symmetry-classification} that this is only the case if the couplings are diagonal in either the spin or valley sector. That is why, in this work, we always consider couplings diagonal in the spin sector as stated in Eq.~\eqref{eq:coupling-matrices}. In the context of moir\'{e} materials, most physically relevant spin-valley models are indeed of this form. This additional assumption, therefore, leaves our method still generally applicable to most models of interest.

In the original implementation of the pf-FRG \cite{reuther2010} and most works since then the flow equations are truncated using the Katanin truncation scheme \cite{katanin2004}, which we also adapt here\footnote{
More recently, an alternative multi-loop truncation has been introduced in the context of electronic FRG calculations \cite{kugler2018},
which was subsequently also adapted in the context of pf-FRG \cite{kiese2021, thoenniss2020}. Such a multi-loop approach can also be applied in the context of spin-valley pf-FRG calculations, but will be left to future exploration.}. In the Katanin truncation only the self-energy $\Sigma^\Lambda$ and the two-particle vertex $\Gamma^\Lambda$ are considered, while higher-order vertex functions are neglected. The flow-equations are then given by
\begin{align}
	\label{eq:self-energy-flow-equation}
    \frac{d}{d\Lambda}\Sigma^\Lambda(1', 1) 
    = - \frac{1}{2\pi} \sum_{2} \Gamma^\Lambda(1', 2, 1, 2) S^\Lambda(\omega_2)
\end{align}
for the self-energy and 
\begin{equation}
	\label{eq:vertex-flow-equations}
	\begin{aligned}
    \frac{d}{d\Lambda} \Gamma^\Lambda(1', 2', 1, 2)  \\
    = -\frac{1}{2\pi} \sum_{3, 4} \bigg[
    &\Gamma^\Lambda(1', 2', 3, 4)\Gamma^\Lambda(3, 4, 1, 2) \\
    -\  &\Gamma^\Lambda(1', 4, 1, 3)\Gamma^\Lambda(3, 2', 4, 2) - (3 \leftrightarrow 4)\\
    +\ & \Gamma^\Lambda(2', 4, 1, 3)\Gamma^\Lambda(3, 1', 4, 2) + (3 \leftrightarrow 4)
    \bigg] \\
    \times \ & G^\Lambda(\omega_3)\partial_\Lambda G^\Lambda(\omega_4),
    \end{aligned}
\end{equation}
for the two-particle vertex. Here, the single-scale propagator is defined as $S^{\Lambda} \equiv -\partial_\Lambda G^{\Lambda}|_{\Sigma^{\Lambda} = \text{const.}}$. Note that the flow equations are formulated in the $T \to 0$ limit and the sums should therefore be understood as $\sum_{1} \equiv \sum_{i_1 s_1 l_1} \int d\omega_1$.

\begin{figure*}[t]
    \centering
    \includegraphics{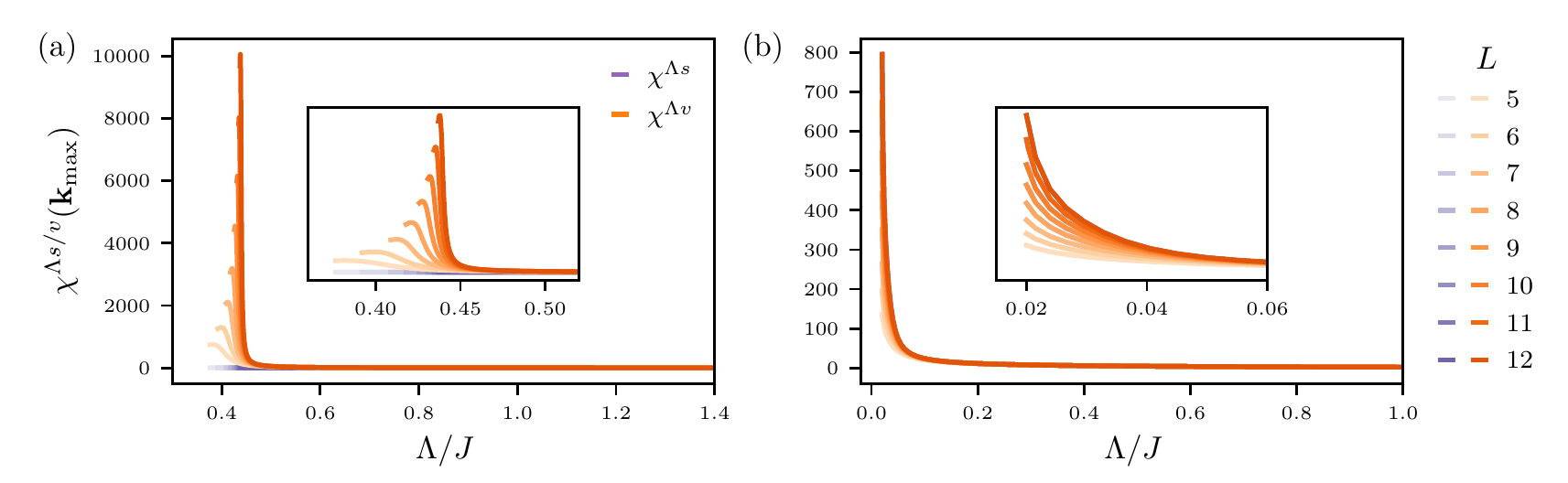}
    \caption{{\bf Flow of the spin and valley structure factor} in a magnetically ordered phase (a) and a paramaganetic phase (b) for different values of the vertex truncation length $L$. All structure factors are shown at the momentum at which they are maximal. The insets zoom into the flow at small cutoffs. In the magnetically ordered phase we clearly see a breakdown of the flow in the valley sector, which manifests as a peak for small $L$ and a more clear divergence when increasing $L$. In the paramagnetic phase the flow is smooth and convex down to about $\Lambda/J = 0.02$, which is the smallest scale for which our calculations are numerically reliable.}
    \label{fig:finite-size-divergence}
\end{figure*}

In order to identify the ground state of a model of interest, we numerically solve the flow equations (as discussed in more detail below in Section \ref{sec:numerics}) and thereby calculate the flow of various 
correlation functions from the flow of the vertex functions. In its most general form, we define a spin-valley-spin-valley correlation function
    \begin{equation}
    \label{eq: magnetic susceptibility}
    \chi_{ij}^{\mu\nu\kappa\lambda}(\omega) = \int_0^\infty d\tau e^{i\omega \tau} \left\langle T_\tau (\sigma^\mu_i \otimes \tau_i^\kappa)(\tau)(\sigma^\nu_j \otimes \tau_j^\lambda)(0)\right\rangle,
\end{equation}
where $T_{\tau}$ is the time-ordering operator. From this general definition we can then read off the form of spin-spin correlations 
\begin{equation}
\label{eq:spin-spin-correlation}
        \chi_{ij}^{s, \mu\nu} \equiv
        \chi_{ij}^{\mu\nu 00} \sim 
        \left\langle \sigma_i^\mu \sigma_j^\nu \right\rangle
\end{equation}
as well as valley-valley correlations
\begin{equation}
\label{eq:valley-valley-correlations}
    \chi_{ij}^{v, \kappa\lambda} \equiv
    \chi_{ij}^{00\kappa\lambda} \sim 
    \left\langle \tau_i^\kappa \tau_j^\lambda \right\rangle.
\end{equation}
A thermal phase transition to long range, symmetry-breaking order in the spin or valley sector at some finite temperature can formally be detected by a divergence in the RG flow of the corresponding correlation at some breakdown scale $\Lambda_c$ \cite{buessen2018a}, as shown in Fig.~\ref{fig:finite-size-divergence}(a). Due to finite numerical resolution, however, they often manifest as a kink or a peak in the susceptibility. The momentum space profile of the dominant structure factor close to the breakdown scale, i.e. the Fourier transform of the static correlation $\chi^{\Lambda s/v}_{ij}(\omega = 0)$, then indicates the type of symmetry-breaking. Since the solution of the flow equation below the breakdown scale $\Lambda_c$ is no longer physical, this only allows us to detect the phase transition that occurs at the largest breakdown scale if there are multiple subsequent transitions. This might be the case when spin and valley degrees of freedom exhibit different ordering transitions at two distinct energy scales. If, in this scenario, the spin sector orders at the larger of the two energy scales, we can not directly determine the ground-state order of the valley sector from the flow of the valley-valley correlations. Instead we need to fall back to, for instance, mean-field arguments as proposed in \cite{kiese2020a} to determine the most likely valley order. If, on the other hand, the correlations show {\sl no} flow breakdown, both spin and valley degrees of freedom do not order, indicative of a ground state that remains paramagnetic or exhibits spin-valley liquid behavior.

These two scenarios are illustrated in Fig.~\ref{fig:finite-size-divergence}(a) and (b). Both panels show the flow of the structure factor at the dominant momentum for a magnetically ordered phase with dominant valley order (a) and the paramagnetic state at the SU(4) point (b) where the spin-valley Hamiltonian corresponds to Eq.~\eqref{eq:su4-hamiltonian}. In the magnetically ordered phase of panel (a) we see a clear flow breakdown in the valley structure factor $\chi^{\Lambda v}$, which manifests as a peak or divergence, depending on the vertex truncation length $L$ (further discussed in Sec.~\ref{sec:numerics}). The spin structure factor $\chi^{\Lambda s}$ shown by the purple lines is strongly suppressed. At the SU(4) point, on the other hand, the flow of the structure factor is smooth and convex down to the lowest energy scale we can reliable calculate ($\Lambda = 0.02 J$), indicating a paramagnetic ground state. Here spin and valley correlations are identical due to the global SU(4) symmetry of the Hamiltonian (and indistinguishable in our plot).

\subsection{Vertex parametrization and symmetry constraints}
\label{sec:vertex-parametrization}
In order to make the solution of the flow equations numerically feasible, one needs to keep the overall number of differential equations 
needed to capture the flow equations as small as possible. Practically, this can be achieved by eliminating redundant calculations
through implementing the symmetry constraints which the Hamiltonian poses on the self-energy and the two-particle vertex. 
A comprehensive symmetry analysis of this sort has been carried out for generic $\mathfrak{su}(2)$ spin models \cite{buessen2019}, which here
will be generalized to the spin-valley Hamiltonians of interest. Details of this symmetry analysis will be discussed in Sec.~\ref{sec:symmetry-classification}, while we will report its main findings in the following. 

The first important finding is that symmetries dictate that the self-energy is completely diagonal and can be parametrized by a single function $\Sigma(\omega)$ as
\begin{equation}
\label{eq:parametrization-self-energy}
	\Sigma(1', 1) = \Sigma(\omega) \delta_{s' s} \delta_{l' l} \delta_{i' i} \delta_{\omega' \omega}.
\end{equation}
We emphasize again that this is only the case if the interactions remain diagonal in either the spin or valley sector. For Hamiltonians with off-diagonal interactions in both sectors the self-energy will not be diagonal in the spin and valley indices, greatly increasing the numerical cost for the solution of the flow equations. The two-particle vertex can be parametrized as
\begin{equation}
	\label{eq:parametrization-two-partile-vertex}
    \begin{aligned}
        	&\Gamma(1', 2', 1, 2) \\
    	&= \Big[ 
    	\Gamma^{\mu\kappa\lambda}_{i_1 i_2}(s, t, u)\ 
    	\theta_{s_{1'}s_1}^\mu \theta_{s_{2'}s_2}^\mu
    	\theta_{l_{1'}l_1}^\kappa \theta_{l_{2'}l_2}^\lambda \  
    	\delta_{i_{1'}i_1} \delta_{i_{2'}i_2} \\
    	&- (1' \leftrightarrow 2') \Big]\  
    	\delta_{\omega_{1'} + \omega_{2'} - \omega_1 - \omega_{2}},
    \end{aligned}
\end{equation}
with the three bosonic transfer frequencies
\begin{equation}
    \label{eq:transfer-frequencies}
    \begin{aligned}
    	s &=\omega_{1'}+\omega_{2'} \\
    	t &=\omega_{1'}-\omega_{1} \\
    	u &=\omega_{1'}-\omega_{2}. 
	\end{aligned}
\end{equation}
This parametrization is of the same form as for $\mathfrak{su}(2)$ spin models -- apart from an increased number of components due to the valley sector $\sim \theta_{l_{1'}l_1}^\kappa \theta_{l_{2'}l_2}^\lambda$ with the corresponding indices $\kappa$ and $\lambda$. If we assume the Hamiltonian to be diagonal in the spin sector, we will only need to consider components diagonal in the spin $\sim \theta_{s_{1'}s_1}^\mu \theta_{s_{2'}s_2}^\mu$, with the corresponding index $\mu$ (and vice-versa for a system with a diagonal valley Hamiltonian). 
The basis functions of the parametrization are constrained by the symmetries of the Hamiltonian as  
\begin{equation}
    \label{eq:self-energy-symmetry-constraints}
    \begin{aligned}
		\Sigma(\omega) &\in i\mathds{R} \\
		\Sigma(\omega) &= -\Sigma(-\omega)
    \end{aligned}
\end{equation}
\begin{equation}
    \label{eq:vertex-symmetry-constraints}
    \begin{aligned}
    	\Gamma^{\mu\kappa\lambda}_{i_1 i_2}(s, t, u) &\in 
    	\begin{cases}
    		\mathds{R}  & \text{if } \xi(\kappa) \xi(\lambda) = 1\\
    		i\mathds{R}  & \text{if } \xi(\kappa) \xi(\lambda) = -1\\
    	\end{cases}\\
    	\Gamma^{\mu\kappa\lambda}_{i_1 i_2}(s, t, u) &= \Gamma^{\mu\lambda\kappa}_{i_2 i_1}(-s, t, u)\\
    	\Gamma^{\mu\kappa\lambda}_{i_1 i_2}(s, t, u) &= \xi(\kappa) \xi(\lambda) \Gamma^{\mu\kappa\lambda}_{i_1 i_2}(s, -t, u)\\
    	\Gamma^{\mu\kappa\lambda}_{i_1 i_2}(s, t, u) &= \xi(\kappa) \xi(\lambda)  \Gamma^{\mu\lambda\kappa}_{i_2 i_1}(s, t, -u)
    \end{aligned}
\end{equation}
where we defined the sign function
\begin{equation}
    \label{eq:xi-of-mu}
	\xi(\kappa) = 
	\begin{cases}
		\ 1 & \text{if } \kappa = 0\\
		-1  & \text{otherwise}
	\end{cases} \ .
\end{equation}
These are the same relations as for the $\mathfrak{su}(2)$ case, apart from a missing constraint relating the $s$ und $u$ frequencies in the two-particle vertex (c.f. Eq.~(14) in Ref.~\cite{buessen2019}). This is a consequence of the Hamiltonian only being invariant under a {\sl global} particle-hole symmetry instead of the {\sl local} particle-hole symmetry under which the $\mathfrak{su}(2)$ Hamiltonian is invariant. We discuss this in more detail in Sec.~\ref{sec:symmetry-classification}.
The missing relation, however, does not change the key implications of the constraints, namely that the basis functions are either completely real or imaginary, and that values of the vertex functions at negative transfer frequencies can be inferred from the positive frequency axes.

The parametrization of the two-particle vertex using the three transfer frequencies in Eq.~\eqref{eq:transfer-frequencies} is convenient for deriving the flow equations and symmetry constraints. However, to better capture the asymptotic frequency dependence of the two-particle vertex one can further refine the frequency parametrization \cite{wentzell2020,kiese2021, thoenniss2020}.
The first step is to group the contributions in the flow-equation of the two-particle vertex given in Eq.~\eqref{eq:vertex-flow-equations} into three channels according to their two-particle irreducibility. This results in a particle-particle ($pp$), direct particle-hole $(dph)$ and crossed particle-hole $(cph)$ channel, which correspond to the three contributions on the right-hand side (RHS) of Eq.~\eqref{eq:vertex-flow-equations}, in the respective ordering. In these terms, the flow equation for the two-particle vertex can be written as 
\begin{equation}
    \frac{d}{d \Lambda} \Gamma^{\Lambda}
	=\dot{g}_{p p}^{\Lambda}
	+\dot{g}_{d p h}^{\Lambda}
	+\dot{g}_{c p h}^{\Lambda}.
\end{equation}
and the vertex is parametrized (stating only the frequency dependence) as
\begin{equation}
\label{eq:channel-parametrization}
	\Gamma^\Lambda(s, t, u) = \Gamma^{\Lambda \to \infty} + \sum_c g_c^\Lambda(\omega_c, v_c, v'_c),
\end{equation}
where $\Gamma^{\Lambda \to \infty}$ is the bare two-particle vertex at infinite cutoff. Each channel $g_c(\omega_c, v_c, v^\prime_c)$ is parametrized by one bosonic transfer frequency $\omega_c$ and two additional fermionic frequencies $v^{\phantom{\prime}}_c, v^\prime_c$. The precise definition of the frequencies can be chosen in numerous ways. It is, however, advantageous to choose them so that the symmetry constraints of the two-particle vertex given in Eq.~\eqref{eq:vertex-symmetry-constraints} result in equally simple relations for each channel in the new parametrization. Here, we adapt the choice of Ref.~\cite{kiese2021}
\begin{equation}
\label{eq:asymptotic-frequencies}
    \begin{aligned}
        \omega_{pp}  &= s  & v_{pp}  &= \omega_1 - \frac{s}{2} & v'_{pp}  &= \frac{s}{2} - \omega_{1'} \\
	    \omega_{dph} &= t  & v_{dph} &= \omega_1 + \frac{t}{2} & v'_{dph} &= \omega_{1'} - \frac{t}{2}  \\
	    \omega_{cph} &= u  & v_{cph} &= \omega_1 - \frac{u}{2} & v'_{cph} &= \omega_{1'} - \frac{u}{2},
    \end{aligned}
\end{equation}
and give the resulting symmetry constraints for the channels in \ref{app:asymptotic-frequency-parametrization}. Compared to $\mathfrak{su}(2)$ spin models, no constraint relating the particle-particle and crossed particle-hole channel with each other is present, which can be traced back to the missing symmetry constraint relating the $s$ and $u$ frequency dependence\footnote{Fortunately, as we will explain in Sec.~\ref{sec:numerics}, this only results in an increase of numerical complexity by a factor of two, making numerical calculations only slightly more costly.}.

To complete the discussion, we still need to state the initial conditions of the flow equations corresponding to the self-energy and two-particle vertex in the limit $\Lambda \to \infty$, which are given by
\begin{equation}
    \label{eq:initial-conditions}
    \begin{aligned}
    	\Sigma^{\Lambda\to\infty}(\omega) &= 0  \\
    	\Gamma_{i_1i_2}^{\Lambda\to\infty \mu\kappa\lambda}(s, t, u)& = \frac{1}{8}J_{s,i_1i_2}^\mu  J_{v, i_1i_2}^{\kappa\lambda},
    \end{aligned}
\end{equation}
with the couplings $J^\mu_{s,i_1i_2}$ and $J^{\kappa\lambda}_{i_1i_2}$ defined in Eq.~\eqref{eq:coupling-matrices}.
\subsection{Numerical implementation} 
\label{sec:numerics}
The numerical solution of the pf-FRG flow equations poses several challenges and necessitates further approximations to be made. To overcome these challenges, we employ the state of the art numerical implementation of Refs.~\cite{kiese2021, thoenniss2020}, where additional details of the implementation are discussed. Here, we only give a short overview and discuss some slight technical differences in the implementation for spin-valley models.

First, one has to truncate the infinite lattice geometry by a finite lattice graph. 
Employing the symmetries of the lattice geometry for which the spin-valley model is formulated and the local U(1) symmetry present in all pseudo-fermion Hamiltonians, the spatial dependence of the two-particle vertex can be reduced to just one site index $j$ and one arbitrary fixed reference site $i_0$, as will be derived in Sec.~\ref{sec:symmetry-classification}. To obtain a finite number of vertex components $\Gamma^\Lambda_{i_0 j}$ (considering only the lattice site dependence), we define a finite length scale $L$ and truncate the vertex $\Gamma^\Lambda_{i_0, j}$ for bond distances $d(i_0, j) > L$, effectively enforcing a maximal correlation length. The finite-size effect of this truncation can be observed in Fig.~\ref{fig:finite-size-divergence}, where several calculations with increasing values of $L$ were performed for a magnetically ordered and a paramagnetic phase. In the ordered phase the flow breakdown sharpens from a relatively broad peak for low values of $L$ to a clear divergence for larger values of $L$, which is a typical observation. The paramagnetic phase is, in contrast, not affected by the increase of $L$ (at least qualitatively).
From an algorithmic point of view, the asymptotic scaling of the computation time is quadratic in the number of lattice points $N_L \sim L^d$, where $d$ is the number of spatial dimensions. This is due to the fact that the number of vertex components as well as the sum over all lattice sites included in the flow equations scale linearly with $N_L$. In this work, we typically perform calculations at $L=9$, above which the breakdown scale does not significantly change anymore and the numerical effort is still reasonable.

Since the pf-FRG approach is formulated at zero temperature, another point we need to address is how to discretize the continuous Matsubara frequencies. To accurately resolve all features of the two-particle vertex, it turns out that particular care needs to be taken in the choice of frequency meshes \cite{kiese2021, thoenniss2020}. To this end, the frequencies are discretized on adaptive, hybrid linear-logarithmic meshes, which are updated using a scanning routine between each step of the ordinary differential equation (ODE) solver. In addition to continuous Matsubara frequencies, the flow equations at $T = 0$ include frequency integrals which have to be performed numerically. To calculate these integrals we employ an adaptive quadrature which takes both the relevant features around the origin and the algebraic decay for large frequencies into account. Values of the vertex for frequencies not lying on the discrete frequency meshes are obtained by multi-linear interpolation.
The computation time asymptotically scales with the number of (positive) bosonic frequencies $N_\Omega$ and (positive) fermionic frequencies $N_\nu$ as $\mathcal{O}(N_\Omega \cdot N_\nu^2)$. A typical set-up for which the two-particle vertex is sufficiently well resolved is $N_\Omega = 40$ and $N_\nu=30$, which we use for all calculations in this work. The computational effort to compute the self-energy is, compared to the vertex, negligible, as it only depends on one frequency. Here we choose a frequency mesh with $N_\Sigma = 250$ frequencies. 
In the $\mathfrak{su}(2)$ case only positive frequencies where required, as the symmetry constraints map all negative frequency components to some positive counterpart. For spin-valley models, however, due to the missing symmetry constraint relating the particle-particle and crossed particle-hole channel (discussed in Sec.~\ref{sec:vertex-parametrization}), we have to also consider negative frequencies for either $\nu_c$ or $\nu'_c$. This results in an additional factor of two in computation time compared to $\mathfrak{su}(2)$ spin models. 

The adaptive frequency meshes and integration routine allow for an efficient evaluation of the RHS of the flow equations. For the solution of the ODEs themselves we choose the Bogacki-Shampine method \cite{bogacki1989}, which is a third-order Runge-Kutta method with adaptive step size control. We find that this method is a good compromise between computational cost and numerical precision.

Although the asymptotic scaling of the computation time with the number of lattice points and frequencies is the same as for the $\mathfrak{su}(2)$ case, more complex spin-valley models usually require a much larger numerical effort, as the extra valley index greatly increases the number of independent two-particle vertex components $\Gamma^{\Lambda, \mu \kappa\lambda}_{i_1i_2}$, in which the computation time scales linearly. With the coupling matrices given in Eq.~\eqref{eq:coupling-matrices}, there would be $N_\Gamma = 4 \cdot 4^2 = 64$ independent vertex components (only considering the spin-valley dependence). In comparison, the parametrization for generic $\mathfrak{su}(2)$ models only has $N_\Gamma = 4^2 = 16$ components. Fortunately, in almost all physical models extra symmetries in the spin and valley space will greatly reduce the number of independent components. Considering, e.g., an SU(2) symmetry in the spin space and a U(1) symmetry in valley space, which is present in several models for moir\'{e} materials \cite{venderbos2018, zhang2019}, the number is already reduced to $N_\Gamma = 2 \cdot 6 = 12$. For these models the numerical effort is similar to $\mathfrak{su}(2)$ models with off-diagonal interactions and even allows for computations of relatively large phase diagrams as will be presented in Sec.~\ref{sec:results}.

%%%%%%%%%%%%%%%%%%%%%%%%%%%%%%%%%%%%%%%%%%%%%%%%%%%%%%%%%%

\section{Symmetry classification}
\label{sec:symmetry-classification}

To proof the validity of the parametrization and the symmetry constraints presented in the previous section, we repeat the symmetry analysis of Ref.~\cite{buessen2019}, where the pseudo-fermion Hamitonian for $\mathfrak{su}(2)$ spin models with generic diagonal and off-diagonal interactions is considered, but for the spin-valley Hamiltonian given in Eq.~\eqref{eq:hamiltonian}. We show that most of the symmetries of the $\mathfrak{su}(2)$ pseudo-fermion Hamiltonian are either also present in the spin-valley Hamiltonian, or can be generalized in a straightforward fashion. There are, however, some differences that we will highlight in the following. Most notably, we show that, even at the SU(4) point, the spin-valley model does not posses a {\sl local} particle-hole symmetry that is present in the $\mathfrak{su}(2)$ case, but only the corresponding {\sl global} symmetry. Consequently, it is also not present in generalizations of the SU(2) Heisenberg model to SU(N), which might not have been clearly stated before. This is the reason for the missing symmetry constraint for the two-particle vertex as presented in the previous section.
\subsection{Local U(1) symmetry}
\label{sec:local-u1-symmetry}
The first symmetry transformation we consider, a local U(1) transformation, directly follows from the form of the spin-valley operator given by Eq.~\eqref{eq:spin-valley-operators}. It  acts on the fermionic Hilbert space at site $i$ by multiplying a local phase $\varphi_i \in [0, 2\pi)$ to the fermionic operators as
\begin{equation}
\label{eq:u(1)-symmetry}
	g_{\varphi_i}
	\begin{pmatrix} f^\dagger_{isl}  \\ 	f^{\pdagger}_{isl} \end{pmatrix} 
	g^{-1}_{\varphi_i} = 
	\begin{pmatrix} e^{i\varphi_i} f^\dagger_{isl} \\  e^{-i\varphi_i} f^{\pdagger}_{isl} \end{pmatrix},
\end{equation}
which clearly leaves all spin-valley operators invariant. Interpreting the spin-valley Hamiltonian as a fermionic representation of an $\mathfrak{su}(4)$ spin model, it is simply a consequence of the choice for the fermionic representation of the spin operators. It is therefore also present in all conventional pf-FRG implementations using the standard pseudo-fermion representation. In that sense, it is sometimes also referred to as a gauge redundancy instead of a symmetry, as it is not a symmetry of the original spin Hamiltonian, but only of the pseudo-fermion representation.
For our functional renormalization group approach we are interested in the implication of the symmetry on the functional form\footnote{Note that our definition deviates from normal ordering to be in line with the conventional definition of retarded Greens functions.} of the one-particle correlation function
\begin{equation}
	\label{eq:one-particle-correlator}
	\begin{aligned}
	G(1', 1) &\equiv - \langle f^\pdagger_{1'}f^{\dagger}_{1}\rangle \\
	&= - \int d \tau' d \tau e^{i \tau' \omega'-i \tau \omega}\left\langle f_{i' \tau' s' l'}^{\pdagger} f^{\dagger}_{i \tau s l}\right\rangle
	\end{aligned}
\end{equation}
and the two-particle correlation function
\begin{equation}
	\label{eq:two-particle-correlator}
	\begin{aligned}
		&G(1', 2', 1, 2) := \langle f^\pdagger_{1'} f^\pdagger_{2'} f^{\dagger}_{2} f^{\dagger}_{1} \rangle  \\
    	&= \int d \tau_{1'} d \tau_{2'} d \tau_{1} d \tau_{2} e^{i(\tau_{1'} \omega_{1'}+\tau_{2'} \omega_{2'}-\tau_{1} \omega_{1}-\tau_{2} \omega_{2})} \\
        &\times \left\langle f_{i_{1'} \tau_{1'} s_{1'} l_{1'}}^{\pdagger} 
        f_{i_{2'} \tau_{2'} s_{2'} l_{2'}}^{\pdagger} 
        f^{\dagger}_{i_{2} \tau_{2} s_2 l_2} 
        f^{\dagger}_{i_{1} \tau_{1} s_1 l_1}\right\rangle,
	\end{aligned}
\end{equation}
where we suppress the time-ordering operator as it becomes trivial in the path integral framework that the function renormalization group is formulated in. Acting with the local U(1) transformation given in Eq.~\eqref{eq:u(1)-symmetry} on the definition of the correlation functions and demanding their invariance leads to the corresponding symmetry constraint. It directly implies that we can restrict ourselves to a local one-particle correlation function
 \begin{equation}
 \label{eq:locality-one-particle-correlator}
	 G(1', 1) = G(1', 1) \delta_{i_{1'}i_1},
 \end{equation}
 which only depends on one lattice site $i_1$, and a bi-local two-particle correlation function
\begin{equation}
    \label{eq:locality-two-particle-correlator}
    \begin{aligned}
        G(1', 2', 1, 2) =\ &G(1', 2', 1, 2) \delta_{i_{1'}i_1} \delta_{i_{2'}i_2} \\
	    -\ &G(2', 1', 1, 2) \delta_{i_{2'}i_1} \delta_{i_{1'}i_2},
    \end{aligned}
\end{equation}
which only depends on the two lattices sites $i_1$ and $i_2$.
\subsection{Global particle-hole symmetry}
\label{sec:particle-hole-symmetry}
In the pf-FRG for $\mathfrak{su}(2)$ spin models spin operators $S_i^a$ are represented using fermions with one spin index $\alpha = \pm 1$ as 
\begin{equation}
    S^a = \frac{1}{2}f^\dagger_{i \alpha} \theta^a_{\alpha \alpha'} f^{\pdagger}_{i \alpha'},
\end{equation}
with $a \in \{1, 2, 3\}$. Additional to the U(1) gauge redundancy, there exists another redundancy in this representation that can be formulated as a {\sl local} particle-hole symmetry \cite{buessen2019}. It acts on the fermionic Hilbert space as
\begin{equation}
	g_i \begin{pmatrix}
		f_{i \alpha}^{\dagger} \\
		f_{i \alpha}^{\pdagger}
    \end{pmatrix} g_i^{-1}=
    \begin{pmatrix}
		\alpha f_{i \bar{\alpha}}^{\pdagger} \\
		\alpha f_{i \bar{\alpha}}^{\dagger}
	\end{pmatrix},
\end{equation}
 with $\bar{\alpha}\equiv-\alpha$. It leaves the fermionic representation of the $\mathfrak{su}(2)$ spin operators invariant and is therefore a symmetry of the pseudo-fermion Hamiltonian. We note that this symmetry is not anti-unitary and therefore does not correspond to the usual physical particle-hole symmetry \cite{buessen2019}. Instead, it is again a consequence of the representation of the spin operators. The natural extension for spin-valley models with spin index $s = \pm 1$ and valley index $l = \pm 1$ is the transformation
\begin{equation}
	g_i \begin{pmatrix} f^\dagger_{isl}  \\ 	f_{isl}^{\pdagger} \end{pmatrix}  g_i^{-1} = 
	\begin{pmatrix} slf_{i\bar{s}\bar{l}}^{\pdagger}\\ slf^\dagger_{i\bar{s}\bar{l}} \end{pmatrix},
\end{equation}
under which the spin-valley operator transforms as 
\begin{equation}
    \label{eq:ph-on-sv-operator}
	g_i\  \sigma_i^\mu \otimes \tau_i^\kappa \ g_i^{-1} = - \xi(\mu) \xi(\kappa) \sigma_i^\mu \otimes \tau_i^\kappa,
\end{equation}
which can be shown straightforwardly using the anti-commutation relations of the fermionic operators and the identity 
\begin{equation}
	\label{eq:pauli-xi-relation}
	\bar\alpha \bar\alpha ' \theta^\mu_{\alpha \alpha'}= \xi(\mu) \theta^\mu_{\bar\alpha ' \bar\alpha}.
\end{equation}
Spin-valley operators with either the spin index $\mu$ or the valley index $\kappa$ set to zero -- which correspond to the individual spin and valley operators as defined in Eq.~\eqref{eq:spin-valley-operators} -- are invariant under this transformation. General spin-valley operators, on the other hand, may change their sign. The Hamiltonian is, therefore, not invariant under the local particle-hole symmetry that acts on the Hilbert space of just one lattice site. Spin-valley operators, however, only appear in pairs in the spin-valley Hamiltonian. Performing the local particle-hole symmetry transformation on {\sl all} lattice sites, such a pair of spin-valley operators transform as
\begin{equation}
    \begin{aligned}
	    g (\sigma_i^\mu \otimes \tau_i^\kappa)  (\sigma_j^\nu \otimes \sigma_j^\lambda) g^{-1} &= \xi(\mu)\xi(\kappa)\xi(\nu)\xi(\lambda)\\
	    &\times (\sigma_i^\mu \otimes \tau_i^\kappa)  (\sigma_j^\nu \otimes \sigma_j^\lambda).
    \end{aligned}
\end{equation}
If an odd number of spin and valley indices is set to zero, this again implies a sign change. Recalling the definition of the spin-valley Hamiltonian in Eq.~\eqref{eq:hamiltonian} and the following definition of the exchange matrices in Eq.~\eqref{eq:coupling-matrices}, such terms are not included in our definition of the Hamiltonian. All terms that do appear in the Hamiltonian are indeed invariant. The main difference to the $\mathfrak{su}(2)$ pseudo-fermion Hamiltonian is, therefore, that the spin-valley Hamiltonian is invariant only under the \textit{global} transformation, while the former was invariant under the \textit{local} transformation.
For the local single-particle correlation function, the global particle-hole symmetry implies 
\begin{equation}
    G(1', 1) \delta_{i' i}= -s s' l l' G(i-\omega \bar{s}\bar{l}, i -\omega' \bar{s}'\bar{l}') \delta_{i' i}
\end{equation}
and for the bi-local two-particle correlator it implies
\begin{equation}
    \begin{aligned}
    	G(1', 2', 1, 2)  \delta_{i_{1'}i_1} \delta_{i_{2'}i_2} 
    	&= s_{1'}s_1l_{1'}l_1 s_{2'}s_2l_{2'}l_2\delta_{i_{1'}i_1} \delta_{i_{2'}i_2} \\
    	& \times G(i_1-\omega_1 \bar{s}_1\bar{l}_1, i_2 - \omega_2 \bar{s}_2\bar{l}_2, \\
    	&i_1-\omega_{1'} \bar{s}_{1'}\bar{l}_{1'}, 
    	i_2 -\omega_{2'} \bar{s}_{2'}\bar{l}_{2'}).
    \end{aligned}
\end{equation}
These relations are, apart form the extra factors of valley indices, the same as for the $\mathfrak{su}(2)$ case when considering the global transformation. The invariance under the local transformation would yield additional constraints on the two-particle correlator acting only on multi-indices with the same lattice site ($i_1$ or $i_2$). For the parametrized two-particle vertex these result in a constraint relating the $s$ and $u$ dependence or, in the asymptotic frequency parametrization defined in Eqs.~(\ref{eq:channel-parametrization}, \ref{eq:asymptotic-frequencies}), the particle-particle and crossed particle-hole channel with each other. As already discussed in Sec.~\ref{sec:overview} this constraint is, consequently, missing for spin-valley models.
\subsection{Generalized time-reversal symmetry}
For $\mathfrak{su}(2)$ spin models, a genuinely physical symmetry is the invariance under time-reversal.  In this setting, time-reversal reverses the sign of all spin operators $S^a \to - S^a$ and, as it is an anti-unitary symmetry, additionally applies complex conjugation to all complex numbers. Hamiltonians with real couplings in which spin operators only appear in pairs are therefore always invariant under time reversal. On the Hilbert space of the $\mathfrak{su}(2)$ pseudo-fermions it can be represented as
\begin{equation}
	g\begin{pmatrix}
		f_{i \alpha}^{\dagger} \\
		f_{i \alpha}^{\pdagger}
	\end{pmatrix} g^{-1}=\begin{pmatrix}
		e^{i \pi \alpha / 2} f_{i \bar{\alpha}}^{\dagger} \\
		e^{-i \pi \alpha / 2} f_{i \bar{\alpha}}^{\pdagger}
	\end{pmatrix}.
\end{equation}
We again consider a straightforward generalization of the transformation to spin-valley operators, which we define as the anti-unitary transformation
\begin{equation}
	g \begin{pmatrix} f^\dagger_{isl}  \\ 	
	f_{isl}^{\pdagger}  \end{pmatrix} g^{-1} = 
	\begin{pmatrix} e^{i\pi s/2}e^{i\pi l/2} f^\dagger_{i\bar{s}\bar{l}} \\ e^{-i\pi s/2}e^{-i\pi l/2} f^{\pdagger}_{i \bar{s}\bar{l}} \end{pmatrix}.
\end{equation}
Using the relation $e^{i\pi(\alpha - \alpha')/2} = \alpha \alpha'$ 
and Eq.~\eqref{eq:pauli-xi-relation}, it is straightforward to show that the spin-valley operator transforms as
\begin{equation}
	 g\  \sigma_i^\mu \otimes \tau_i^\kappa \ g^{-1} = \xi(\mu) \xi(\kappa) \sigma_i^\mu \otimes \tau_i^\kappa,
\end{equation} 
which, up to a minus sign, is the same transformation behavior as for the particle-hole symmetry in Eq.~\eqref{eq:ph-on-sv-operator}. As only pairs of spin-valley operators appear in the spin-valley Hamiltonian, for which the minus sign is irrelevant, the arguments for the invariance of Hamiltonian given there, consequently, also apply here.
Applying this generalized version of time-reversal to the local one-particle correlator implies
\begin{equation}
	G(1', 1) \delta_{i', i} = s s' l l' G(i-\omega'\bar{s}'\bar{l}', i -\omega \bar{s}\bar{l})^*\delta_{i', i}
\end{equation}
where the complex conjugation stems from the fact that the transformation is anti-unitary. For the bi-local two-particle correlation function it implies
\begin{equation}
	\begin{aligned}
	    G(1', 2', 1, 2)  \delta_{i_{1'}i_1} \delta_{i_{2'}i_2} &= s_{1'}s_1l_{1'}l_1 s_{2'}s_2l_{2'}l_2 \delta_{i_{1'}i_1} \delta_{i_{2'}i_2} \\
	    &\times G(i_1-\omega_{1'}\bar{s}_{1'}\bar{l}_{1'}, i_2 -\omega_{2'} \bar{s}_{2'}\bar{l}_{2'}, \\
	    &i_1 -\omega_1 \bar{s}_1\bar{l}_1, i_2 -\omega_2 \bar{s}_2\bar{l}_2)^*.
    \end{aligned}
\end{equation}
Apart from extra valley indices, this is exactly the same as in the $\mathfrak{su}(2)$ case.
\subsection{Hermitian symmetry}
Just as the $\mathfrak{su}(2)$ spin operator the spin-valley operator is Hermitian. The spin-valley Hamiltonian only consists of pairs of spin-valley operators and we have restricted ourselves to real couplings, making it Hermitian aswell. Complex transposition therefore leaves the Boltzman factor in the thermal expectation value invariant. Applying complex transposition on both sides of Eqs.~(\ref{eq:one-particle-correlator}, \ref{eq:two-particle-correlator}) and explicitly evaluating the RHS by ``pulling'' the complex transpose into the thermal expectation value, we obtain the constraint  
\begin{equation}
	G(1', 1)\delta_{i', i} = G(i -\omega s l, i - \omega' s' l')^* \delta_{i', i}
\end{equation}
for the local one-particle correlator and 
\begin{equation}
	\begin{aligned}
	    G(1', 2', 1, 2)  \delta_{i_{1'}i_1} \delta_{i_{2'}i_2} &=  \delta_{i_{1'}i_1} \delta_{i_{2'}i_2} \\
	    &\times G(i_{1} -\omega_1 s_1 l_2, i_2 -\omega_2 s_2 l_2,\\
	    &i_1 - \omega_{1'} s_{1'} l_{1'}, i_2 -\omega_{2'} s_{2'} l_{2'})^* 
	\end{aligned}
\end{equation}
for the two-particle correlator. These constraints are again of the same form as for the $\mathfrak{su}(2)$ case. 
\subsection{Lattice symmetries}
\label{sec:lattice-symmetries}
The spin models we consider are all formulated on lattices that can be specified in terms of an underlying Bravais lattice and a possibly multi-atomic basis. Therefore, lattice symmetries exist necessarily for any spin-valley model and are very important to efficiently implement the pf-FRG. Their implementation is the same whether one considers $\mathfrak{su}(2)$ spin models or spin-valley models. We can therefore use the same approach as for the conventional pf-FRG as, e.g., explained in Ref.~\cite{buessen2019}. There, all sites are assumed to be identical, in the sense that one can map any site to any other site via a lattice automorphism $T$ that leaves the lattice itself invariant. On the fermionic operators, such a transformation acts as
\begin{equation}
	g_{T}\begin{pmatrix}
		f_{i sl}^{\dagger} \\
		f_{i sl}^{\pdagger} 
	\end{pmatrix} g_{T}^{-1}
	=\begin{pmatrix}
		f_{T(i) sl}^{\dagger} \\
		f_{T(i) sl}^{\pdagger} 
	\end{pmatrix}.
\end{equation}
In the case of bond-directional couplings, the transformation would additionally have to be combined with transformations in spin and valley space. For the one-particle correlation function this implies 
 \begin{equation}
 	G\left(1' , 1\right)\delta_{i', i}=G\left(T\left(i\right) \omega' s'l', T(i) \omega s l \right)\delta_{i', i}.
 \end{equation}
  The locality constraint in Eq.~\eqref{eq:locality-one-particle-correlator}, resulting from the local U(1) symmetry, already reduces the spatial dependence of the self-energy to only one site index $i_1$. Using lattice automorphisms, we can map all sites to an arbitrary reference site $i_0$ and therefore completely remove the spatial dependence of the one-particle correlation function.
 Similarly, for the two-particle correlation function it implies
 \begin{equation}
    \begin{aligned}
		&G\left(1', 2' , 1,2\right)\delta_{i_{1'}i_1} \delta_{i_{2'}i_2} = \delta_{i_{1'}i_1} \delta_{i_{2'}i_2}\\
        &\times G\big(T\left(i_{1}\right) \omega_{1'} s_{1'} l_{1'}, T\left(i_{2}\right) \omega_{2'} s_{2'}l_{2'}, \\
        & T\left(i_{1}\right) \omega_{1} s_1l_1, T\left(i_{2}\right) \omega_{2} s_2 l_2 \big).
	\end{aligned}
\end{equation}
Combining this with the bi-locality constraint in Eq.~\eqref{eq:locality-two-particle-correlator}, and again mapping the first index $i_1$ to an arbitrary reference site $i_0$, the spatial dependence of the two-particle correlator can be reduced to just one lattice site.
\subsection{Parametrization of correlation functions}
In order to make use of the symmetry constraints on the correlation functions it is advantageous to parametrize them so that the symmetry constraints manifest in a more practical form. To this end, we can extent the parametrization for the correlation functions for generic $\mathfrak{su}(2)$ spin models introduced in \cite{buessen2019} also to spin-valley models. This ultimately leads to the parametrization of the self-energy and two-particle vertex in Eqs.~(\ref{eq:parametrization-self-energy}, \ref{eq:parametrization-two-partile-vertex}) and the symmetry constraints in Eqs.~(\ref{eq:self-energy-symmetry-constraints}, \ref{eq:vertex-symmetry-constraints}). Starting with the one-particle correlation function, we argued that due to the local U(1) symmetry and lattice symmetries it is independent of the lattice site. Additionally, due to Matsubara frequency conservation, which is a consequence of translational invariance in imaginary time, it is diagonal in the frequency arguments. Expanding the spin and valley dependence in Pauli matrices $\theta^\mu \theta^\kappa$ ($\mu,\kappa = 0, 1, 2, 3$), the one-particle correlation function can be parametrized as
\begin{equation}
	\label{eq:parametrization-1p-correlator}
		G(1', 1) = G^{\mu\kappa}(w) \theta_{s' s}^\mu \theta_{l' l}^\kappa \delta_{i'i} \delta_{\omega' \omega}.
\end{equation}
Similarly, the two-particle correlation function depends only on two lattice sites and three frequencies, for which we choose the three transfer frequencies defined in Eq.~\eqref{eq:transfer-frequencies}. The parametrization then reads 
\begin{equation}
	\label{eq:parametrization-2p-correlator}
	\begin{aligned}
		&G(1', 2', 1, 2) 
		\\&= \Big(G^{\mu\nu\kappa\lambda}_{i_1 i_2}(s, t, u) \theta_{s_{1'}s_1}^\mu \theta_{s_{2'}s_2}^\nu \theta_{l_{1'}l_1}^\kappa \theta_{l_{2'}l_2}^\lambda\delta_{i_{1'}i_1} \delta_{i_{2'}i_2} \\
		& - (1' \leftrightarrow 2')\Big)  
		\delta_{\omega_{1'} + \omega_{2'}-\omega_1 -\omega_{2}}.
	\end{aligned}
\end{equation}
Plugging this parametrization into the symmetry constraints derived in Secs.~\ref{sec:local-u1-symmetry}-\ref{sec:lattice-symmetries} we obtain the symmetry constraints for the basis functions of the parametrization listed in Table.~\ref{tab:symmtry-constraints-correlators}. In the derivation of these constraints we make heavy use of Eq.~\eqref{eq:pauli-xi-relation} and the particle exchange symmetry
\begin{equation}
   G(1', 2', 1, 2) = G(2', 1', 2, 1)
\end{equation}
which is present in all purely fermionic models. 
\begin{table}[h]
	\begin{align*}	
			G^{\mu\kappa}(\omega) &= \xi(\mu)\xi(\kappa) G^{\mu\kappa}(\omega) & (\text{H} \circ \text{TR})\\
			G^{\mu\kappa}(\omega) &= -G^{\mu\kappa}(-\omega)                & (\text{H} \circ \text{TR} \circ \text{PH} )\\
			G^{\mu\kappa}(\omega) &= -G^{\mu\kappa}(\omega)^*               & (\text{TR}\circ \text{PH}) \\
			G^{\mu\nu\kappa\lambda}_{i_1 i_2}(s, t, u) &= \xi(\mu)\xi(\nu)\xi(\kappa)\xi(\lambda) \\
			&\times G^{\mu\nu\kappa\lambda}_{i_1 i_2}(s, t, u)^*	
			&(\text{TR} \circ \text{PH} \circ \text{H} \circ \text{TR})\\
			G^{\mu\nu\kappa\lambda}_{i_1 i_2}(s, t, u) &= G^{\nu\mu\lambda\kappa}_{i_2 i_1}(-s, t, u) 
		    &(\text{H} \circ \text{TR} \circ \text{PH} \circ \text{X}) \\
			G^{\mu\nu\kappa\lambda}_{i_1 i_2}(s, t, u) &= \xi(\mu)\xi(\nu)\xi(\kappa)\xi(\lambda) \\
			&\times G^{\mu\nu\kappa\lambda}_{i_1 i_2}(s, -t, u)
			&(\text{H} \circ \text{TR})\\
			G^{\mu\nu\kappa\lambda}_{i_1 i_2}(s, t, u) &= \xi(\mu)\xi(\nu)\xi(\kappa)\xi(\lambda) \\
			&\times G^{\nu\mu\lambda\kappa}_{i_2 i_1}(s, t, -u) 
			&(\text{H} \circ \text{TR} \circ \text{X})
	\end{align*}
	\caption{Symmetry constraints for the basis functions of the parametrization of the correlation functions. The labels specify which symmetries of the Hamiltonian were used in their derivation, where H stand for Hermitian, TR for generalized time-reversal, PH for global particle-hole and X for particle-exchange symmetry. The most notable implications are that all correlation functions will always be either only real or imaginary and all expression with negative frequencies can be related to those with positive frequencies.}
	\label{tab:symmtry-constraints-correlators}
\end{table}

The list of symmetry constraints is very similar to the $\mathfrak{su}(2)$ case derived in Ref.~\cite{buessen2019}, but has two significant differences. Firstly, as already discussed in Secs.~\ref{sec:vertex-parametrization} and \ref{sec:particle-hole-symmetry}, the symmetry constraint relating $s$ and $u$ frequencies, or the particle-particle and crossed particle-hole channel, is missing because the spin-valley Hamiltonian is not invariant under a \textit{local} particle-hole transformation but only under the \textit{global} version. Secondly, the symmetry constraints do not imply that the one-particle correlation function is completely diagonal in all spin and valley indices. In the parametrization this would manifest in $G^{00}$ being the only non-vanishing basis function. Instead, for a general spin-valley Hamiltonian, also  the terms $G^{ab}$ with $a, b > 0$, which come with the factor $\sim\theta^a_{ss'}\theta^b_{ll'}$, are allowed. This would increase the number of flow equations and therefore also the numerical complexity significantly. Additionally, we could not use the conventional pf-FRG flow equations given in Eqs.~(\ref{eq:self-energy-flow-equation}, \ref{eq:vertex-flow-equations}), where a diagonal one particle-correlator (and self-energy) was assumed. Fortunately, in the context of moir\'{e} materials, many Hamiltonians of physical relevance posses additional symmetries in the spin and valley space \cite{venderbos2018, zhang2019} that further constrain the spin and valley dependence of the self-energy and two-particle vertex. It turns out that the minimal symmetry needed in order for the one-particle correlator to be diagonal is a $\mathds{Z}_2 \times \mathds{Z}_2 \times \mathds{Z}_2$ symmetry in either the spin or valley sector. On the level of spin-valley operators this means that the Hamiltonian is invariant under the transformation (for the case of the spin sector)
\begin{equation}
    \label{eq:Z2-symmetry}
	g_\mu\  \sigma_i^\mu \otimes \tau_i^\kappa \ g_\mu^{-1} = \xi(\mu) \sigma_i^\mu \otimes \tau_i^\kappa,
\end{equation}
for each $\mu$ individually. This simply reverses the signs of all $\sigma^\mu_i \otimes \tau^\kappa_i$ with $\mu > 0$. Assuming a completely diagonal spin exchange matrix as in Eq.~\eqref{eq:coupling-matrices}, the spin-valley Hamiltonian is indeed invariant under this transformation. This directly implies that all terms proportional to a single $\sim \theta^\mu$ (with $\mu>0$) in the correlation functions have to vanish. More precisely, it imposes the constraint
\begin{equation}
    G^{\mu\kappa}(\omega) = \delta_{\mu 0} G^{0\kappa}(\omega),
\end{equation}
which in combination with the first equation in Table.~\ref{tab:symmtry-constraints-correlators} implies
\begin{equation}
    G^{\mu\kappa}(\omega) = \delta_{\mu 0}\delta_{\kappa0} G^{00}(\omega) \equiv \delta_{\mu 0}\delta_{\kappa0} G(\omega),
\end{equation}
resulting in a completely diagonal one-particle correlation function parametrized by a single basis function $G(\omega)$. For the coupling matrices stated in Eq.~\eqref{eq:coupling-matrices}, we can therefore use the standard pf-FRG appraoch also for spin-valley models.
Assuming this additional symmetry, in the two-particle correlator only diagonal components in the spin sector $\sim \theta^\mu \theta^\mu$ (no sum over $\mu$) are allowed, resulting in the constraint
\begin{equation}
    G^{\mu\nu\kappa\lambda}_{i_1 i_2}(s, t, u) = \delta_{\mu\nu} G^{\mu\mu\kappa\lambda}_{i_1 i_2}(s, t, u) \equiv  \delta_{\mu\nu} G^{\mu\kappa\lambda}_{i_1 i_2}(s, t, u).
\end{equation}
Imposing these additional constraints, all factors of $\xi(\mu)\xi(\nu)$ in Table.~\ref{tab:symmtry-constraints-correlators} are equal to one and the relations reduce exactly to the constraints given in Eqs.~(\ref{eq:self-energy-symmetry-constraints}, \ref{eq:vertex-symmetry-constraints}) with the self-energy and two-particle vertex replaced by the one- and two-particle correlation functions. We can therefore still consider a completely imaginary one-particle correlator that is odd in frequency space and completely diagonal.
The two-particle correlator is either completely real or imaginary, depending on the sign of $\xi(\kappa, \lambda)$, and all negative frequency components can be mapped to a positive counterpart.

The argument why these constraints on the disconnected correlation functions carry over to the one-particle irreducible correlation functions, i.e. the self-energy and the vertex, is the same as given for the $\mathfrak{su}(2)$ case in \cite{buessen2019}. For the self-energy it simply follows from the Dyson equation \cite{kopietz2010}
\begin{equation}
	 G(1', 1) = \frac{1}{i\omega - \Sigma(1', 1)},
\end{equation}
from which it is easy to see that all constraints carry over to the self-energy.
For the two-particle vertex the tree expansion (neglecting the three-particle vertex) relates it to the connected two-particle correlation function $G^{(c)}$ as \cite{kopietz2010}
\begin{equation}
\begin{aligned}
    	&G^{(c)}\left(1', 2' , 1,2\right)= \\
    	&-\sum_{3,4,5,6} \Gamma(3,4 , 5,6) G\left(1' , 3\right) G\left(2' , 4\right) G(5 , 1) G(6 , 2).
\end{aligned}
\end{equation}
As the one-particle correlation function is diagonal in all indices, it is clear that all constraints carry over from the connected correlation function to the two-particle vertex. That the constraints from the disconnected two-particle correlation function carry over to the connected correlation function can be proven by their definition via generating functionals \cite{kopietz2010}.
\subsection{Symmetries of the flow equations}
\label{sec:flow-equation-symmetries}
To verify that the parametrization and the symmetry constraints derived in the previous sections are indeed preserved also for the flowing self-energy and two-particle vertex for any value of $\Lambda$, they can additionally be proven using the pf-FRG flow equations given in Eqs.~(\ref{eq:self-energy-flow-equation}, \ref{eq:vertex-flow-equations}) . That the parametrization for the self-energy in Eq.~\eqref{eq:parametrization-self-energy} and for the two-particle vertex in Eq.~\eqref{eq:parametrization-two-partile-vertex} is indeed complete can be seen by inserting them into the RHS of the flow equations and confirming that no additional terms are generated.

For the additional symmetry constraints the proof can be performed via induction, as already explained in Refs.~\cite{reuther, buessen2019}. This essentially amounts to verifying the fulfillment of the constraints in the initial conditions and then showing that the derivatives $\frac{d}{d\Lambda}\Sigma$ and $\frac{d}{d\Lambda}\Gamma$ given by the RHS of the flow equations also fulfill them, assuming the self-energy and two-particle vertex themselves already do. The proof that the self-energy is odd, imaginary and completely diagonal has to be repeated for spin-valley models due to slight differences in the flow equations. This is quite lengthy and, therefore, done in \ref{app:flow-equation-symmetries}. For the two-particle vertex, the proof of the symmetry constraints is much easier on the level of the unparametrized vertex, as there the flow equations still have a much simpler form. We therefore postulate the relations
\begin{align}
	\Gamma^\Lambda(1', 2', 1, 2) &= \Gamma^\Lambda(2', 1', 2, 1)\label{eq:particle-exchange}\\
	\Gamma^\Lambda(1', 2', 1, 2) &= \Gamma^\Lambda(1, 2, 1', 2')^* \label{eq:full-vertex-conjugate}\\
	\Gamma^\Lambda(1', 2', 1, 2) &= \Gamma^\Lambda(-2', -1', -2, -1) \label{eq:full-vertex-exchange}\\
	\Gamma^\Lambda(1', 2', 1, 2) &= s_{1'}s_1l_{1'}l_1 s_{2'}s_2l_{2'}l_2\nonumber\\ &\times \Gamma^\Lambda(\bar{1}, \bar{2}, \bar{1}', \bar{2}') \label{eq:vertex-H-TR},
\end{align}
where we defined $-1 = (i_1 -\omega_1 s_1 l_1)$ and $\bar{1} = (i_1 \omega_1 \bar{s}_1 \bar{l}_1)$.
 When translated to the parametrized two-particle vertex and then combined, these relations yield exactly the symmetry constraints given in Eq.~\eqref{eq:vertex-symmetry-constraints}. Proving the relations for the unparametrized vertex, therefore, directly proves the symmetry constraints of the parametrized vertex. As Eq.~\eqref{eq:particle-exchange} simply amounts to a simple particle exchange, no further proof is required. Eq.~\eqref{eq:full-vertex-conjugate} is proven in \cite{buessen2019} and Eq.~\eqref{eq:full-vertex-exchange} in \cite{reuther} using the general pf-FRG flow equations. The only remaining relation still left to prove is Eq.~\eqref{eq:vertex-H-TR}, which we also show in \ref{app:flow-equation-symmetries}. This proves that the parametrization and the symmetry constraints are indeed valid also for the flowing self-energy and vertex, at any value of the cutoff $\Lambda$. 

%%%%%%%%%%%%%%%%%%%%%%%%%%%%%%%%%%%%%%%%%%%%%%%%%%%%%%%%%%

%\clearpage
\section{Results}
\label{sec:results}
To give an explicit example for the application of the pseudo-fermion functional renormalization group approach introduced in the manuscript and its efficient implementation in terms of the aforementioned symmetries, we apply it to elucidate the phase diagram of an SU(2)$_\mathrm{spin}$ $\otimes$ U(1)$_\mathrm{valley}$ symmetric spin-valley Hamiltonian on the triangular lattice. The explicit Hamiltonian we consider is 
\begin{equation}
\label{eq:su2xu1-hamiltonian}
    \begin{aligned}
        \mathcal{H} &= \frac{J}{8} \sum_{\langle ij\rangle} 
        (1 + \boldsymbol{\sigma}_i\boldsymbol{\sigma}_j)(1 + \boldsymbol{\tau}_i \boldsymbol{\tau}_j) \\
    	&+ \frac{J_{x}}{8} \sum_{\langle i j \rangle} 
    	(1 + \boldsymbol{\sigma}_i\boldsymbol{\sigma}_j)(\tau_i^x \tau_j^x + \tau_i^y \tau_j^y)\\
        &+ \frac{J_{z}}{8} \sum_{\langle i j \rangle} 
    	(1 + \boldsymbol{\sigma}_i\boldsymbol{\sigma}_j)(\tau_i^z \tau_j^z),
    \end{aligned}
\end{equation}
with a SU(4) symmetric term proportional to the coupling $J$ and an in-plane $J_x$ and out-of-plane $J_z$ coupling that when non-zero break the SU(4) symmetry down to an SU(2) symmetry in the spin sector and a U(1) symmetry in the valley sector. We only include interactions between nearest neighbours $\langle ij \rangle$.

Such a model can be motivated, e.g., from including the effect of Hund's type couplings in a two-orbital extended Hubbard model and performing a strong coupling expansion \cite{natori2019}. It can therefore be regarded as a natural extension to previously studied models with either full SU(4) or reduced SU(2)$_\mathrm{spin}$ $\otimes$ SU(2)$_\mathrm{valley}$ symmetry \cite{kiese2020a, corboz2012, natori2018, natori2019} by adding an XXZ type perturbation to the orbital sector and likewise provides an intermediate, but important step towards the more complicated spin-valley Hamiltonians proposed for various moir\'e systems \cite{xian2019, natori2019, zhang2019}.

\subsection{Phase diagram}
\begin{figure}
    \centering
    \includegraphics{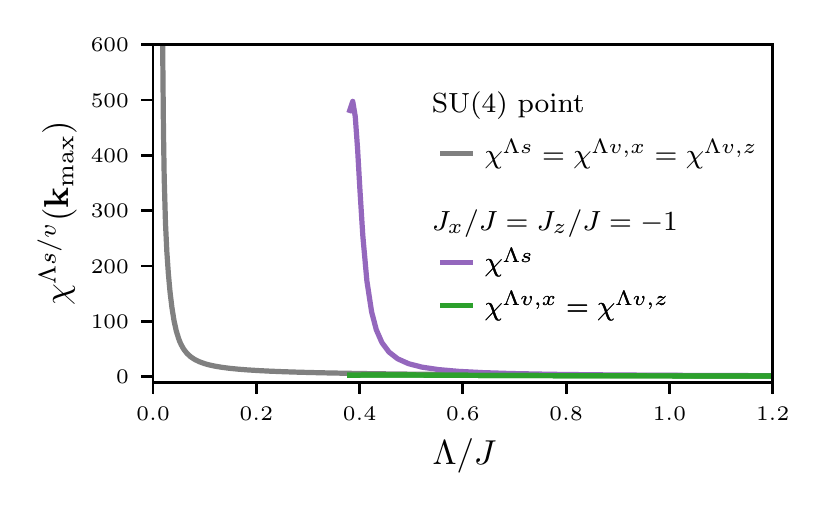}
    \caption{{\bf Flow of the structure factor at points of higher symmetry}. All structure factors are shown at the momentum where they are maximal. The grey line shows the structure factor at the SU(4) point, where the considered spin-valley model corresponds to the SU(4) symmetric Heisenberg model. Here, all structure factor components are identical. The flow is smooth and convex down to the lowest numerically reliable cutoff and no flow breakdown occurs, indicating a putative quantum spin-valley liquid (QSVL) ground state. The purple and green lines show the spin and valley structure factor for $J_x/J = J_z/J = 0$, where all terms containing valley operators cancel and the spin-valley model resembles an SU(2) symmetric Heisenberg model. In this case, the valley structure factors are strongly suppressed and the spin structure factor shows a sharp peak at the $K$ and $K'$ points, indicating $120^\circ$ order in the spin sector. }
    \label{fig:su4-flow}
\end{figure}
\begin{figure}
    \centering
    \includegraphics[width = 0.45\textwidth]{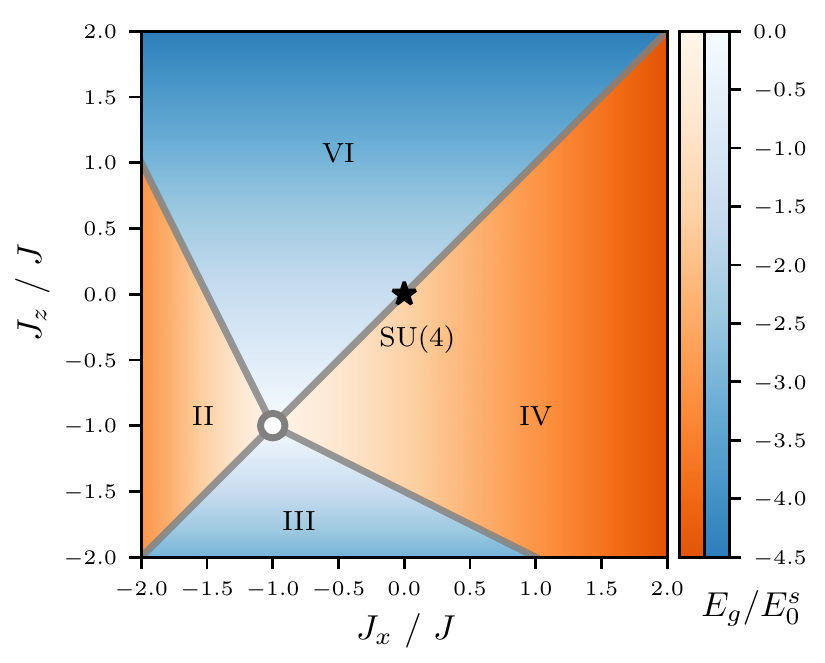}
    \caption{{\bf Classical phase diagram in valley space for fixed spin ordering} obtained from a Luttinger-Tisza analysis. The grey lines depict the phase boundaries and the color illustrates the (normalized) ground state energy, where blue denotes out-of-plane and orange in-plane ordering. At $J_x/J = J_z/J = - 1$, where the phase boundaries meat, the classical mean-field Hamiltonian vanishes. Away from this point the Luttinger-Tisza analysis predicts the following types of valley order: (II) in-plane ferromagnetic (FM), (III) out-of-plane FM, (IV) in-plane $120^\circ$, (VI) out-of-plane $120^\circ$. The so obtained valley order is independent from the fixed nearest-neighbour spin order.}
    \label{fig:classical-phasediagram}
\end{figure}
\begin{figure*}
    \centering
    \includegraphics{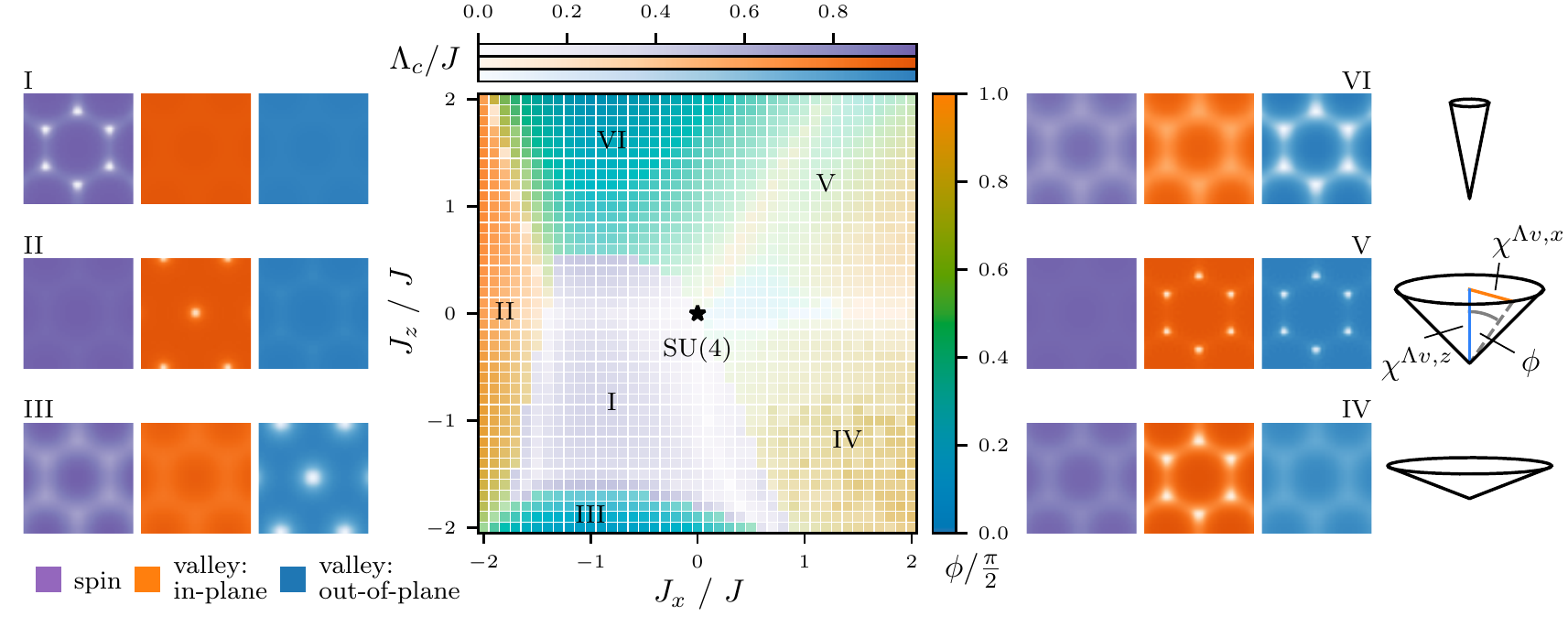}
    \caption{{\bf Phase diagram of the SU(2)$_\mathrm{spin}$ $\otimes$ U(1)$_\mathrm{valley}$ symmetric spin-valley model.} The color indicates which structure factor is dominant at the breakdown scale $\Lambda_c$, where purple implies dominant spin order, colors between orange and yellow indicate dominant order in the valley sector and the opacity determines the magnitude of the breakdown scale $\Lambda_c/J$. In the case where we observe a flow breakdown in both the in-plane ($\chi^{\Lambda v, x}$) and out-of-plane ($\chi^{\Lambda v, x}$) valley structure factor the color determines the angle $\phi$ illustrated in the cones on the right of the figure. (I-VI) show the structure factors at $\Lambda_c$ for the different types of order we observe: (I) 120$^\circ$ spin order, (II) out-of-plane FM valley order,  (III) in-plane FM valley order, (IV-VI) 120$^\circ$ valley order shifting from an out-of-plane (IV) to an in-plane (VI) orientation, with competing order (V) in between. For $J_x/J = J_z/J = 0$, indicated by the star, the model is equivalent to the SU(4) symmetric Heisenberg model for which no flow breakdown is observed.}
    \label{fig:quantum-phasediagram}
\end{figure*}
\begin{figure}
    \centering
    \includegraphics{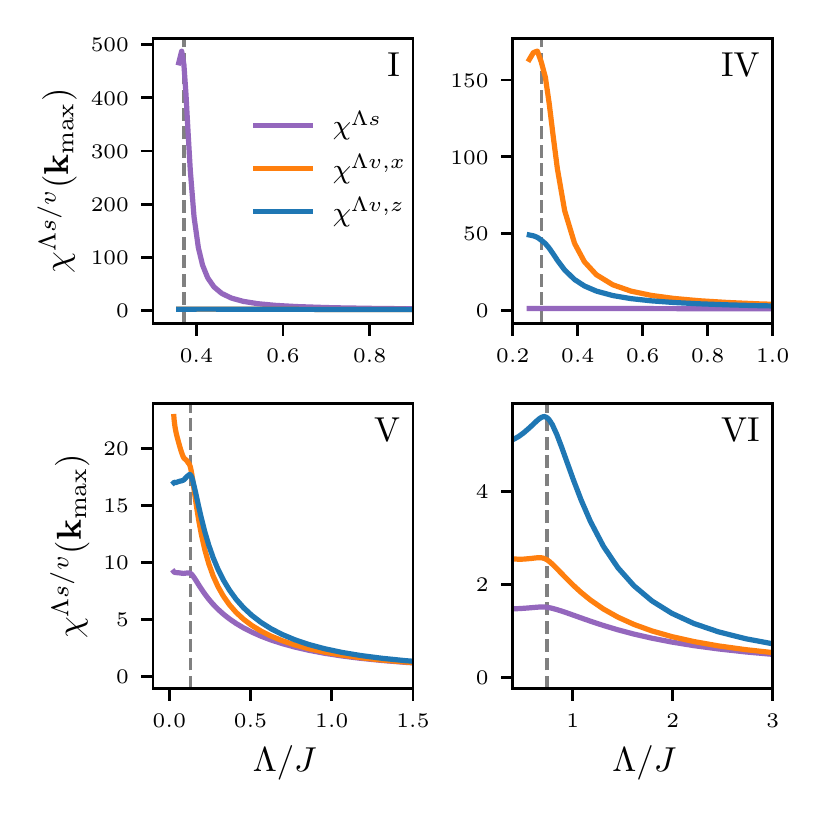}
    \caption{{\bf Flow of the structure factors} for different types of order as described under Fig.~\ref{fig:quantum-phasediagram}. The dashed lines show the breakdown scale $\Lambda_c$. (I) shows dominant spin order (purple) with the valley structure factors strongly suppressed. (IV-VI) shows dominant valley order which shifts from an in-plane (orange) to an out-of-plane (blue) orientation.}
    \label{fig:flows}
\end{figure}
\begin{figure}
    \centering
    \includegraphics{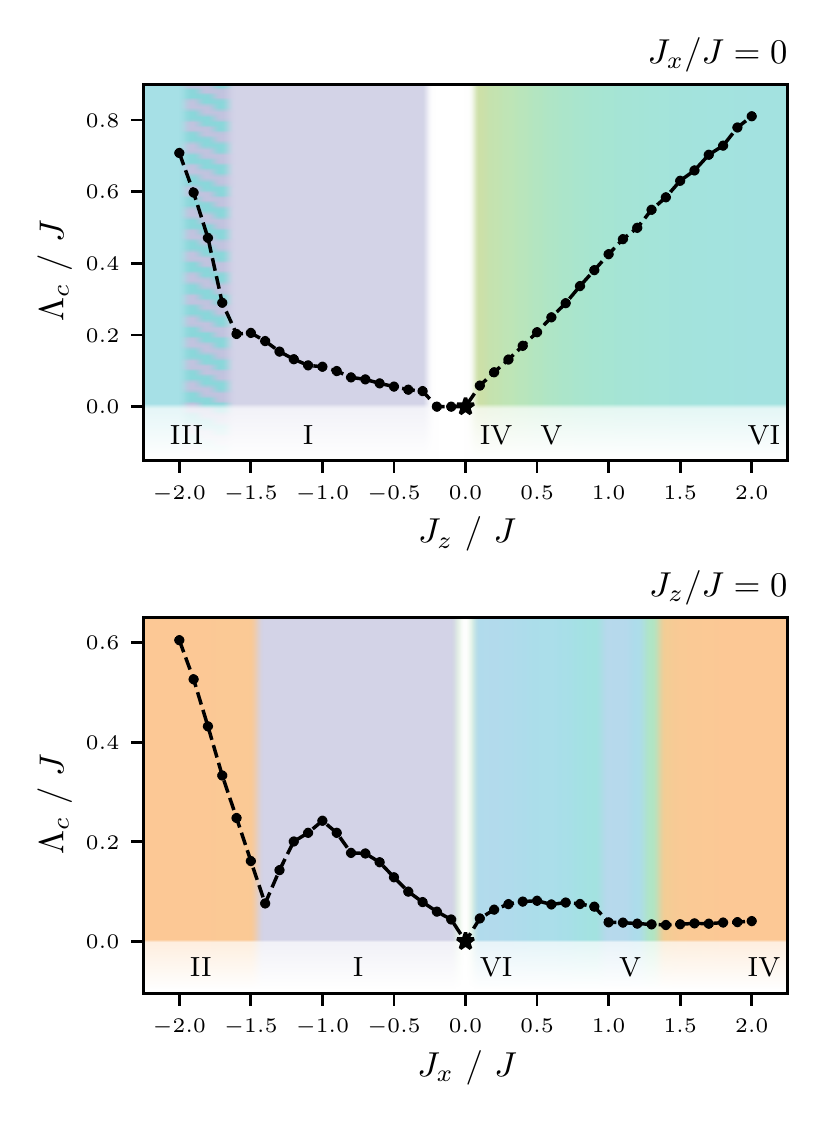}
    \caption{{\bf Cuts through the phase diagram} of Fig.~\ref{fig:quantum-phasediagram} at $J_x = 0$ (top) and $J_z = 0$ (bottom). The color-coding and the labels (I-VI) denote different types of order and are explained in Fig.~\ref{fig:quantum-phasediagram}. 
    In the transition between III and I along the $J_x = 0$ axis, which is colored both blue and purple, both $\chi^{\Lambda s}$ and $\chi^{\Lambda v, z}$ show flow breakdowns at a similar $\Lambda_c$ and with similar magnitudes. The white regimes close to the SU(4) point mark points for which no flow breakdown is observed.}
    \label{fig:zero-lines}
\end{figure}
To obtain the quantum phase diagram we fix the coupling $J$ in front of the SU(4) symmetric part of the Hamiltonian in Eq.~\eqref{eq:su2xu1-hamiltonian} to a positive value and then vary the values of the in-plane coupling $J_x$ and out-of-plane coupling $J_z$ which break the SU(4) symmetry. As described in Sec.~\ref{sec:overview}, to determine the magnetic order for a particular pair of couplings $(J_x, J_z)$ we  calculate the flow of the spin-spin and valley-valley correlations (and  associated structure factors) as defined in Eqs.~(\ref{eq:spin-spin-correlation}, \ref{eq:valley-valley-correlations}), check whether or not a flow breakdown occurs and if so, which type of order is visible in the structure factor at the breakdown scale $\Lambda_c$. Due to the SU(2)$_\mathrm{spin}$ $\otimes$ U(1)$_\mathrm{valley}$ symmetry of the Hamiltonian  all nonvanishing components of the spin-spin correlation are equivalent and we calculate only $\chi^{\Lambda s}_{ij} \equiv \chi^{\Lambda s, xx}_{ij} = \chi^{\Lambda s, yy}_{ij} = \chi^{\Lambda s, zz}_{ij}$. For the valley-valley correlation we can distinguish between in-plane and out-of-plane order by calculating the in-plane valley-valley correlation  $\chi^{\Lambda v, x}_{ij} \equiv \chi^{\Lambda v, xx}_{ij} = \chi^{\Lambda v, yy}_{ij}$ and out-of-plane valley-valley correlation $\chi^{\Lambda v, z}_{ij} \equiv \chi^{\Lambda v, zz}_{ij}$. 

Starting at the SU(4) point with $J_x/J = J_z/J = 0$, where all spin-spin and valley-valley correlations are equivalent, we observe no flow breakdown of the structure factors, as depicted by the grey line in Fig.~\ref{fig:su4-flow}. This indicates that no magnetic order is present in both the spin and the valley sector even for very low energy scales and indicates a putative quantum spin-valley liquid (QSVL) state
\cite{kiese2020a}. Going away from the SU(4) point, however, we almost immediately observe a flow breakdown in either the spin or valley sector, indicating that the putative QSVL state is highly unstable in the presence of XXZ like perturbations. This is in line with results for the $\mathfrak{su}(2)$ XXZ model on the triangular lattice, where by varying the out-of-plane coupling a phase transition from in-plane $120^\circ$ order to an ``umbrella" order is observed at the SU(2) symmetric point \cite{sellmann2015}. Similarly, we observe a rich ensemble off different spin and valley ordered phases with both in- and out-of-plane ordering in the valley sector. 

Before we present the full quantum phase diagram, however, let us first consider a classical mean-field approach to better understand the origin of the different phases. To this end, we note that the spin sector by itself will order either ferromagnetically (FM) or in a $120^\circ$ order, depending on the sign of the exchange coupling. Assuming one of these states is realized, we decouple the spin and valley sector by approximating the pair of spin operators by its expectation value $\boldsymbol{\sigma}_{i}\boldsymbol{\sigma}_{j} \approx \langle \boldsymbol{\sigma}_{i}\boldsymbol{\sigma}_{j}\rangle$, with $\langle \boldsymbol{\sigma}_{i}\boldsymbol{\sigma}_{j}\rangle = 1$ for ferromagnetic (FM) and $\langle \boldsymbol{\sigma}_{i}\boldsymbol{\sigma}_{j}\rangle = \cos(2\pi/3) = -0.5$ for $120^\circ$ order. The resulting mean-field Hamiltonian is then given, up to a constant, by
\begin{equation}
    \begin{aligned}
        &\mathcal{H}_{\mathrm{MF}} = \\
        &E^s_0 \sum_{\langle ij \rangle} \bigg[\left(1 +\frac{J_{x}}{J}\right) \left(\tau_i^x \tau_j^x + \tau_i^y \tau_j^y\right) 
    	+ \left(1 + \frac{J_z}{J}\right) \tau_i^z \tau_j^z \bigg],
    \end{aligned}
\end{equation}
where the spin expectation value only appears in the positive factor $E_0^s \equiv J(1+\langle \boldsymbol{\sigma}_{i}\boldsymbol{\sigma}_{j}\rangle)$ and, therefore, has no influence on the type of valley order. Approximating the valley operators by classical vectors with $\abs{\boldsymbol{\tau}_i}=1$ we perform a Luttinger-Tisza analysis \cite{luttinger1946, luttinger1951} on the mean-field Hamiltonian. This analysis predicts in-plane (out-of-plane) valley order for large values of $\abs{1+J_x/J}$ ($\abs{1+J_z/J}$), which is either FM for positive, or $120^\circ$ like for negative values. The precise phase boundaries along with the ground-state energies $E_g$ are depicted in Fig.~\ref{fig:classical-phasediagram}. 

Special attention needs to be paid to the point at $J_x/J = J_z/J = -1$ where the phase boundaries meet. Exactly at this point, the couplings in front of the valley operators are equal to zero and the mean-field Hamiltonian vanishes. Going back to the full quantum Hamiltonian, it reduces to only the term $\sum_{ij} J (1 + \boldsymbol{\sigma}_i \boldsymbol{\sigma}_{j})$, which resembles an SU(2) symmetric Heisenberg model with antiferromagnetic coupling $J$. Here, the flow of the spin structure factor shows a sharp peak while the valley structure factors are strongly suppressed, as is depicted in Fig.~\ref{fig:su4-flow}. The same behavior occurs in a larger region around $J_x/J = J_z/J = -1$, which is shown in Fig.~\ref{fig:quantum-phasediagram} along with the corresponding momentum resolved structure factor (annotated with the numeral I). The spin structure factor (in purple) shows strong peaks at the $K$ and $K'$ points, while the in-plane (orange) and out-of-plane (blue) valley structure factors show no distinct features when shown on the same color scale. This indicates $120^\circ$ spin order, which again agrees with results for the conventional $\mathfrak{su}(2)$ Heisenberg model \cite{sellmann2015}.

In all other regions of the quantum phase diagram the valley structure factors are clearly dominant and the spin structure factor shows only weak features. We enumerate the different types of valley order we find by the numerals II-VI, as shown in Fig.~\ref{fig:quantum-phasediagram}. The valley order at large negative couplings (II and III) agrees with the classically predicted results, as either the in- or out-of-plane structure factors show strong peaks at the $\Gamma$ point, indicating FM order. At larger positive values for either the in-plane or out-of-plane coupling (VI - IV) the valley structure factors show peaks at the $K$ and $K'$ points indicating $120^\circ$ like order. In contrast to the sharp phase boundary in the classical case, however, the valley order seems to gradually shift from mostly in-plane (IV), over competing in- and out-of-plane (V) to out-of-plane (VI) order. This is well visualized by the flow of the structure factors in Fig.~\ref{fig:flows}. The valley structure factors both show flow breakdowns at approximately the same breakdown scale, but the magnitude at the breakdown scale shift from a dominant $\chi^{\Lambda v, x}$ to a dominant $\chi^{\Lambda v, z}$  when going from IV to VI. To quantify this transition, we define the angle 
\begin{equation}
    \phi = \text{arctan}(\chi^{\Lambda_c v, x}/ \chi^{\Lambda_c v, z}),
\end{equation} illustrated in Fig.~\ref{fig:quantum-phasediagram} by the cones on the right and by the color scale ranging from from blue (in-plane) over green (competing in-plane and out-of-plane) to orange (out-of-plane) 

To better illustrate the transitions between the different types of order, Fig.~\ref{fig:zero-lines} shows the  $J_x = 0$ and $J_z = 0$ axes of the phase diagram, separately. Starting with the $J_x = 0$ axis at negative $J_z/J$, the at first very dominant out-of-plane valley order (III) gradually transitions to dominant spin order (I), with a region in between where the spin and valley structure factors are of similar magnitude. The kink in the breakdown scale, usually indicating a phase transition, appears at the largest $J_z/J$ where the valley structure factor still shows a clear flow breakdown ($J_z/J \approx -1.6$), even though the spin structure factor is already dominant for smaller $J_z/J$. This is similar at all boundaries of phase I, which also becomes evident in the phase diagram of Fig.~\ref{fig:quantum-phasediagram} by noting that the minima of the breakdown scale are positioned slightly inwards in the region of dominant spin order (colored in purple). Going to positive $J_z/J$, the gradual transition from slightly dominant in-plane (IV) to out-of-plane (VI) order is clearly visible. On the $J_z = 0$ axis similar behavior can be observed, although the transitions appear more sharply. For positive $J_x/J$ the breakdown scale stays extremely small ($\Lambda_c/J < 0.1$), making its precise numerical determination difficult. This suggests that fluctuations, which suppress long-range order, are particularly strong close to the transition between in-plane (IV) and competing in- and ouf-of-plane (V) order.
%%%%%%%%%%%%%%%%%%%%%%%%%%%%%%%%%%%%%%%%%%%%%%%%%%%%%%%%%%

\section{Summary}

In this manuscript, we have presented a generalization of the established pf-FRG approach to generic spin-valley Hamiltonians in the self-conjugate representation of $\mathfrak{su}$(4), with either diagonal spin or valley interactions. We performed a careful symmetry analysis and derived a set of constraints on the vertex functions, which drastically lower the computational cost of tracking the flow of running couplings. Using a highly accurate solver for the functional flow equations, we subsequently applied this method to map out the quantum phase diagram of an SU(2)$_{\textrm{spin}}$ $\otimes$ U(1)$_{\textrm{valley}}$ model on the triangular lattice, which presents a simplified variant of the more general Hamiltonian proposed for TLG/h-BN, but already hosts a rich variety of spin and valley ordered ground states. In addition, we were able to demonstrate, that, by promoting the spin symmetry group from SU(2) to SU(4), quantum fluctuations are boosted, ultimately resulting in a smooth RG flow down to the lowest energy scales, indicative of a spin-valley liquid state. However, this QSVL state appears to be very sensitive even to weak XXZ anisotropies in the valley sector and we almost immediately detect the emergence of long-range order, when perturbing it.

While our focus in this manuscript has been on spin-valley Hamiltonians, we note that very similar models have been discussed for spin-orbit coupled systems that go beyond the celebrated Kugel-Khomskii model. 
The microscopic ingredients of such spin-orbital models are surprisingly similar to those of ``Kitaev materials" \cite{trebst2022} -- a partially filled $4d$ or $5d$ orbital, the formation of a spin-orbital entangled local moment, and an edge-sharing octahedral crystalline environment. Specifically, a $d^1$ configuration can lead to local $j=3/2$ moments subject to bond-directional exchanges that break the original SU(4) symmetry of the $j=3/2$ moments. As a concrete material candidate exhibiting this microscopic mechanism, $\alpha$-ZrCl$_3$ -- a 4$d$ sister compound of the isostructural Kitaev material RuCl$_3$ -- has been put forward \cite{yamada2018}. To study the phase diagram of spin-orbital ground states in such a setting with varying diagonal and off-diagonal couplings, on can again rely on the pseudo-fermion FRG approach put forward in this manuscript.

%%%%%%%%%%%%%%%%%%%%%%%%%%%%%%%%%%%%%%%%%%%%%%%%%%%%%%%%%%
\begin{acknowledgements}
We thank F. L. Buessen, Y. Iqbal, and M. Scherer for discussions.
Partial support from the Deutsche Forschungsgemeinschaft (DFG) -- Projektnummer 277146847 -- SFB 1238 (project C02) is gratefully acknowledged.
The numerical simulations were performed on the JURECA Booster and JUWELS cluster at the Forschungszentrum Juelich as well as the 
CHEOPS cluster at RRZK Cologne.
\end{acknowledgements}

\bibliographystyle{spphys}       % APS-like style for physics
\bibliography{pffrg}   % name your BibTeX data base

\begin{thebibliography}{10}
\providecommand{\url}[1]{{#1}}
\providecommand{\urlprefix}{URL }
\expandafter\ifx\csname urlstyle\endcsname\relax
  \providecommand{\doi}[1]{DOI \discretionary{}{}{}#1}\else
  \providecommand{\doi}{DOI \discretionary{}{}{}\begingroup
  \urlstyle{rm}\Url}\fi

\bibitem{cao2018}
Y.~Cao, V.~Fatemi, A.~Demir, S.~Fang, S.L. Tomarken, J.Y. Luo, J.D.
  Sanchez-Yamagishi, K.~Watanabe, T.~Taniguchi, E.~Kaxiras, R.C. Ashoori,
  P.~Jarillo-Herrero, Nature \textbf{556}(7699), 80 (2018).
\newblock \urlprefix\url{https://doi.org/10.1038/nature26154}

\bibitem{chen2020}
C.~Shen, Y.~Chu, Q.~Wu, N.~Li, S.~Wang, Y.~Zhao, J.~Tang, J.~Liu, J.~Tian,
  K.~Watanabe, et~al., Nat. Phys. \textbf{16}(5), 520–525 (2020).
\newblock \urlprefix\url{https://doi.org/10.1038/s41567-020-0825-9}

\bibitem{liu2020}
X.~Liu, Z.~Hao, E.~Khalaf, J.Y. Lee, Y.~Ronen, H.~Yoo, D.~Haei~Najafabadi,
  K.~Watanabe, T.~Taniguchi, A.~Vishwanath, et~al., Nature \textbf{583}(7815),
  221–225 (2020).
\newblock \urlprefix\url{https://doi.org/10.1038/s41586-020-2458-7}

\bibitem{cao2020}
Y.~Cao, D.~Rodan-Legrain, O.~Rubies-Bigorda, J.M. Park, K.~Watanabe,
  T.~Taniguchi, P.~Jarillo-Herrero, Nature \textbf{583}(7815), 215–220
  (2020).
\newblock \urlprefix\url{https://doi.org/10.1038/s41586-020-2260-6}

\bibitem{cao2018a}
Y.~Cao, V.~Fatemi, S.~Fang, K.~Watanabe, T.~Taniguchi, E.~Kaxiras,
  P.~Jarillo-Herrero, Nature \textbf{556}(7699), 43 (2018).
\newblock \urlprefix\url{https://doi.org/10.1038/nature26160}

\bibitem{lu2019}
X.~Lu, P.~Stepanov, W.~Yang, M.~Xie, M.A. Aamir, I.~Das, C.~Urgell,
  K.~Watanabe, T.~Taniguchi, G.~Zhang, A.~Bachtold, A.H. MacDonald, D.K.
  Efetov, Nature \textbf{574}(7780), 653 (2019).
\newblock \urlprefix\url{https://doi.org/10.1038/s41586-019-1695-0}

\bibitem{chen2019}
G.~Chen, A.L. Sharpe, P.~Gallagher, I.T. Rosen, E.J. Fox, L.~Jiang, B.~Lyu,
  H.~Li, K.~Watanabe, T.~Taniguchi, J.~Jung, Z.~Shi, D.~Goldhaber-Gordon,
  Y.~Zhang, F.~Wang, Nature \textbf{572}(7768), 215 (2019).
\newblock \urlprefix\url{https://doi.org/10.1038/s41586-019-1393-y}

\bibitem{sharpe2019}
A.L. Sharpe, E.J. Fox, A.W. Barnard, J.~Finney, K.~Watanabe, T.~Taniguchi, M.A.
  Kastner, D.~Goldhaber-Gordon, Science \textbf{365}(6453), 605 (2019).
\newblock \urlprefix\url{https://doi.org/10.1126/science.aaw3780}

\bibitem{koshino2018}
M.~Koshino, N.F.Q. Yuan, T.~Koretsune, M.~Ochi, K.~Kuroki, L.~Fu, Phys. Rev. X
  \textbf{8}, 031087 (2018).
\newblock \urlprefix\url{https://doi.org/10.1103/PhysRevX.8.031087}

\bibitem{xian2019}
L.~Xian, D.M. Kennes, N.~Tancogne-Dejean, M.~Altarelli, A.~Rubio, Nano Lett.
  \textbf{19}(8), 4934–4940 (2019).
\newblock \urlprefix\url{https://doi.org/10.1021/acs.nanolett.9b00986}

\bibitem{zhang2019}
Y.H. Zhang, T.~Senthil, Phys. Rev. B \textbf{99}, 205150 (2019).
\newblock \urlprefix\url{https://doi.org/10.1103/PhysRevB.99.205150}

\bibitem{xu2018}
C.~Xu, L.~Balents, Phys. Rev. Lett. \textbf{121}(8), 087001 (2018).
\newblock \urlprefix\url{https://doi.org/10.1103/PhysRevLett.121.087001}

\bibitem{yuan2018}
N.F.Q. Yuan, L.~Fu, Phys. Rev. B \textbf{98}, 045103 (2018).
\newblock \urlprefix\url{https://doi.org/10.1103/PhysRevB.98.045103}

\bibitem{po2018}
H.C. Po, L.~Zou, A.~Vishwanath, T.~Senthil, Phys. Rev. X \textbf{8}, 031089
  (2018).
\newblock \urlprefix\url{https://doi.org/10.1103/PhysRevX.8.031089}

\bibitem{classen2019}
L.~Classen, C.~Honerkamp, M.M. Scherer, Phys. Rev. B \textbf{99}(19), 195120
  (2019).
\newblock \urlprefix\url{https://doi.org/10.1103/PhysRevB.99.195120}

\bibitem{kugel1982}
K.I. Kugel, D.I. Khomski{\u{\i}}, Sov. Phys. Usp. \textbf{25}(4), 231 (1982).
\newblock \urlprefix\url{https://doi.org/10.1070/pu1982v025n04abeh004537}

\bibitem{feiner1997}
L.F. Feiner, A.M. Ole\ifmmode~\acute{s}\else \'{s}\fi{}, J.~Zaanen, Phys. Rev.
  Lett. \textbf{78}, 2799 (1997).
\newblock \urlprefix\url{https://doi.org/10.1103/PhysRevLett.78.2799}

\bibitem{corboz2012}
P.~Corboz, M.~Lajk\'o, A.M. L\"auchli, K.~Penc, F.~Mila, Phys. Rev. X
  \textbf{2}, 041013 (2012).
\newblock \urlprefix\url{https://doi.org/10.1103/PhysRevX.2.041013}

\bibitem{natori2018}
W.M.H. Natori, E.C. Andrade, R.G. Pereira, Phys. Rev. B \textbf{98}(19), 195113
  (2018).
\newblock \urlprefix\url{https://doi.org/10.1103/PhysRevB.98.195113}

\bibitem{natori2019}
W.M.H. Natori, R.~Nutakki, R.G. Pereira, E.C. Andrade, Phys. Rev. B
  \textbf{100}(20) (2019).
\newblock \urlprefix\url{https://doi.org/10.1103/PhysRevB.100.205131}

\bibitem{kiese2020a}
D.~Kiese, F.L. Buessen, C.~Hickey, S.~Trebst, M.M. Scherer, Phys. Rev. Res.
  \textbf{2}(1), 013370 (2020).
\newblock \urlprefix\url{https://doi.org/10.1103/PhysRevResearch.2.013370}

\bibitem{reuther2010}
J.~Reuther, P.~W\"olfle, Phys. Rev. B \textbf{81}(14), 144410.
\newblock \urlprefix\url{https://doi.org/10.1103/PhysRevB.81.144410}

\bibitem{thoenniss2020}
J.~Thoenniss, M.K. Ritter, F.B. Kugler, J.~von Delft, M.~Punk,
  \urlprefix\url{http://arxiv.org/abs/2011.01268}

\bibitem{kiese2021}
D.~Kiese, T.~Mueller, Y.~Iqbal, R.~Thomale, S.~Trebst,
  \urlprefix\url{http://arxiv.org/abs/2011.01269}

\bibitem{bistritzer2011}
R.~Bistritzer, A.H. MacDonald, PNAS \textbf{108}(30), 12233 (2011).
\newblock \urlprefix\url{https://doi.org/10.1073/pnas.1108174108}

\bibitem{Khomskii2014}
D.I. Khomskii, \textit{Transition Metal Compounds} (Cambridge University Press,
  2014)

\bibitem{buessen2019}
F.L. Buessen, V.~Noculak, S.~Trebst, J.~Reuther, Phys. Rev. B \textbf{100}(12),
  125164 (2019).
\newblock \urlprefix\url{https://doi.org/10.1103/PhysRevB.100.125164}

\bibitem{buessen2018a}
F.L. Buessen, D.~Roscher, S.~Diehl, S.~Trebst, Phys. Rev. B \textbf{97}(6),
  064415 (2018).
\newblock \urlprefix\url{https://doi.org/10.1103/PhysRevB.97.064415}

\bibitem{Wetterich1993}
C.~Wetterich, Phys. Lett. B \textbf{301}, 90 (1993).
\newblock \urlprefix\url{https://doi.org/10.1016/0370-2693(93)90726-X}

\bibitem{kopietz2010}
P.~Kopietz, L.~Bartosch, F.~Sch{\"{u}}tz, \textit{Introduction to the
  Functional Renormalization Group} (Springer, Berlin, 2010).

\bibitem{katanin2004}
A.A. Katanin, Phys. Rev. B \textbf{70}(11), 115109 (2004).
\newblock \urlprefix\url{https://doi.org/10.1103/PhysRevB.70.115109}

\bibitem{kugler2018}
F.B. Kugler, J.~{von Delft}, Phys. Rev. B \textbf{97}(3), 035162 (2018).
\newblock \urlprefix\url{https://doi.org/10.1103/PhysRevB.97.035162}

\bibitem{wentzell2020}
N.~Wentzell, G.~Li, A.~Tagliavini, C.~Taranto, G.~Rohringer, K.~Held,
  A.~Toschi, S.~Andergassen, Phys. Rev. B \textbf{102}(8), 085106 (2020).
\newblock \urlprefix\url{https://doi.org/10.1103/PhysRevB.102.085106}

\bibitem{bogacki1989}
P.~Bogacki, L.F. Shampine, Appl. Math. Lett. \textbf{2}(4), 321 (1989).
\newblock \urlprefix\url{https://doi.org/10.1016/0893-9659(89)90079-7}

\bibitem{venderbos2018}
J.W.F. Venderbos, R.M. Fernandes, Phys. Rev. B \textbf{98}(24), 245103 (2018).
\newblock \urlprefix\url{https://doi.org/10.1103/PhysRevB.98.245103}

\bibitem{reuther}
J.~Reuther, PhD thesis, Karlsruhe  (2011).
\newblock \urlprefix\url{https://doi.org/10.5445/IR/1000023236}

\bibitem{sellmann2015}
D.~Sellmann, X.F. Zhang, S.~Eggert, Phys. Rev. B \textbf{91}(8), 081104 (2015).
\newblock \urlprefix\url{https://doi.org/10.1103/PhysRevB.91.081104}

\bibitem{luttinger1946}
J.M. Luttinger, L.~Tisza, Phys. Rev. \textbf{70}(11-12), 954 (1946).
\newblock \urlprefix\url{https://doi.org/10.1103/PhysRev.70.954}

\bibitem{luttinger1951}
J.M. Luttinger, Phys. Rev. \textbf{81}(6), 1015 (1951).
\newblock \urlprefix\url{https://doi.org/10.1103/PhysRev.81.1015}

\bibitem{trebst2022}
S.~Trebst, C.~Hickey, Phys. Rep. \textbf{950}, 1 (2022).
\newblock \urlprefix\url{https://doi.org/10.1016/j.physrep.2021.11.003}

\bibitem{yamada2018}
M.~Yamada, M.~Oshikawa, G.~Jackeli, Phys. Rev. Lett. \textbf{121}, 097201
  (2018).
\newblock \urlprefix\url{https://doi.org/10.1103/PhysRevLett.121.097201}

\end{thebibliography}

%%%%%%%%%%%%%%%%%%%%%%%%%%%%%%%%%%%%%%%%%%%%%%%%%%%%%%%%%%

\clearpage
\appendix

%%%%%%%%%%%%%%%%%%%%%%%%%%%%%%%%%%%%%%%%%%%%%%%%%%%%%%%%%%
\onecolumn
\section{Symmetry constraints in the asymptotic frequency parametrization}
\label{app:asymptotic-frequency-parametrization}
In Eq.~\eqref{eq:vertex-symmetry-constraints} we stated the symmetry constraints of the two-particle vertex in the frequency parametrization using the three transfer frequencies $s$, $t$, and $u$. As was discussed in Sec.~\ref{sec:vertex-parametrization}, in our implementation of the pf-FRG we use a refined frequency parametrization \cite{wentzell2020, kiese2021, thoenniss2020}, where the vertex is split into three channels $g_c(\omega_c, v_c, v'_c)$ as defined in Eq.~\eqref{eq:channel-parametrization}, with our choice of frequencies given in Eq.~\eqref{eq:asymptotic-frequencies}. We can obtain symmetry constraints for the different channels by employing the same parametrization in the spin, valley and site indices as for the full vertex
\begin{equation}
\begin{aligned}
	g^\Lambda_c(1', 2', 1, 2) = 
	\left[g^{\Lambda\mu\kappa\lambda}_{c,i_1 i_2}(\omega_c, v_c, v'_c)\ \theta_{s_{1'}s_1}^\mu \theta_{s_{2'}s_2}^\mu \theta_{l_{1'}l_1}^\kappa \theta_{l_{2'}l_2}^\lambda \  \delta_{i_{1'}i_1} \delta_{i_{2'}i_2} 
	- (1' \leftrightarrow 2')\right] \delta_{\omega_{1'} + \omega_{2'}, \omega_1 + \omega_{2}},
\end{aligned}
\end{equation}
and utilizing that the frequencies $\omega_c, v_c, v'_c$ can be written as linear combinations of the transfer frequencies. Combining one or more symmetry constraints of the two-particle vertex, this results in symmetry constraints for the particle-particle channel 
\begin{equation}
\begin{aligned}
	g^{\Lambda  \mu\kappa\lambda}_{pp,i_1 i_2} (s, v_s, v_s') &= g^{\Lambda  \mu\lambda\kappa}_{pp,i_2 i_1} (-s, v_s, v_s') \\
	g^{\Lambda  \mu\kappa\lambda}_{pp,i_1 i_2} (s, v_s, v_s') &= g^{\Lambda  \mu\lambda\kappa}_{pp,i_2 i_1} (s, -v_s, -v_s') \\
	g^{\Lambda  \mu\kappa\lambda}_{pp,i_1 i_2} (s, v_s, v_s') &= \xi(\kappa) \xi(\lambda) g^{\Lambda  ,\mu\lambda\kappa}_{pp,i_2 i_1} (s, v_s', v_s),
\end{aligned}
\end{equation}
the direct particle-hole channel
\begin{equation}
\begin{aligned}
	g^{\Lambda  \mu\kappa\lambda}_{dph,i_1 i_2} (t, v_t, v_t') &= \xi(\kappa) \xi(\lambda) g^{\Lambda  \mu\kappa\lambda}_{dph,i_1 i_2} (-t, v_t, v_t') \\
	g^{\Lambda  \mu\kappa\lambda}_{dph,i_1 i_2} (t, v_t, v_t') &= \xi(\kappa) \xi(\lambda) g^{\Lambda  \mu\kappa\lambda}_{dph,i_1 i_2} (t, -v_t, -v_t')\\
	g^{\Lambda  \mu\kappa\lambda}_{dph,i_1 i_2} (t, v_t, v_t') &= \xi(\kappa) \xi(\lambda) g^{\Lambda  ,\mu\lambda\kappa}_{dph,i_2 i_1} (t, v_t', v_t),
\end{aligned}
\end{equation}
and the crossed particle-hole channel
\begin{equation}
\begin{aligned}
	g^{\Lambda  \mu\kappa\lambda}_{cph,i_1 i_2} (u, v_u, v_u') &= \xi(\kappa) \xi(\lambda) g^{\Lambda  ,\mu\lambda\kappa}_{cph,i_2 i_1} (-u, v_u, v_u')
	\\
	g^{\Lambda  \mu\kappa\lambda}_{cph,i_1 i_2} (u, v_u, v_u') &= \xi(\kappa) \xi(\lambda) g^{\Lambda  ,\mu\lambda\kappa}_{cph,i_2 i_1} (u, -v_u, -v_u') \\
	g^{\Lambda  \mu\kappa\lambda}_{cph,i_1 i_2} (u, v_u, v_u') &= \xi(\kappa) \xi(\lambda) g^{\Lambda  \mu\kappa\lambda}_{cph,i_1 i_2} (u, v_u', v_u),
\end{aligned}
\end{equation}
where for the fermionic frequencies $v_c, v'_c$ we used the subscripts $s, t, u$ instead of $pp, dph, cph$ for brevity. Using these symmetry relations, we only have to explicitly calculate the two-particle vertex for positive values of $\omega_c$ and $v_c$, but have to also consider negative values for $v'_c$. Additionally, we only have to calculate components with $\abs{v'_c} < \abs{v_c}$. We note again that, compared to the $\mathfrak{su}2$ case, no constraints relating the particle-particle and crossed particle-hole channel are  present.

%%%%%%%%%%%%%%%%%%%%%%%%%%%%%%%%%%%%%%%%%%%%%%%%%%%%%%%%%%
\section{Proof of symmetry constraints via flow equations}
\label{app:flow-equation-symmetries}
In Sec.~\ref{sec:flow-equation-symmetries} we claim that the completeness of the parametrization given in Eqs.~(\ref{eq:parametrization-self-energy}, \ref{eq:parametrization-two-partile-vertex}) and the symmetry constraints given in Eqs.~(\ref{eq:self-energy-symmetry-constraints}, \ref{eq:vertex-symmetry-constraints}) can also be proven by induction using the flow equations, as was already done for the $\mathfrak{su}2$ case \cite{reuther, buessen2019}. The proof amounts to checking that the constraints are fulfilled in the initial conditions and then showing that the RHS of the pf-FRG flow equations in Eqs.~(\ref{eq:self-energy-flow-equation}, \ref{eq:vertex-flow-equations}) also fulfill the constraints, assuming the self-energy and two-particle vertex themselves already do. That the constraints are fulfilled in the initial conditions is easy to see, as for $\Lambda\to\infty$ the two-particle vertex is frequency independent and the self-energy vanishes. We will, therefore, only perform the induction step here.

Starting with the two-particle vertex, it is straightforward to see that the parametrization is complete by plugging it in the pf-FRG flow equations and showing that no additional terms are generated. To proof the symmetry constrains, we postulated equivalent constraints for the unparametrized two-particle vertex in Eqs.~(\ref{eq:particle-exchange}-\ref{eq:vertex-H-TR}), which, when combined, lead to the symmetry constraints of the parametrized vertex. Fortunately, only the relation
\begin{equation}
\label{eq:vertex-left-to-proof}
	\Gamma^\Lambda(1', 2', 1, 2) = s_{1'}s_1l_{1'}l_1 s_{2'}s_2l_{2'}l_2 
	\Gamma^\Lambda(\bar{1}, \bar{2}, \bar{1}', \bar{2}')
\end{equation}
differs from the $\mathfrak{su}(2)$ case and all other relations have already been proven \cite{reuther, buessen2019}. The induction step for this relation is performed by writing down the flow equations for 
     $s_{1'}s_1l_{1'}l_1 s_{2'}s_2l_{2'}l_2 \Gamma^\Lambda(\bar{1}, \bar{2}, \bar{1}', \bar{2}')$
and then manipulating the RHS
\begin{equation} 
\begin{aligned}    
 	&s_{1'}s_1l_{1'}l_1 s_{2'}s_2l_{2'}l_2 \frac{d}{d\Lambda} \Gamma^\Lambda(\bar{1}, \bar{2}, \bar{1}', \bar{2}') \nonumber\\
 	&= -s_{1'}s_1l_{1'}l_1 s_{2'}s_2l_{2'}l_2 \frac{1}{2\pi} \sum_{3, 4} \bigg[
 	\Gamma^\Lambda(\bar{1}, \bar{2}, 3, 4)\Gamma^\Lambda(3, 4, \bar{1}', \bar{2}')
 -	\Gamma^\Lambda(\bar{1}, 4, \bar{1}', 3)\Gamma^\Lambda(3, \bar{2}, 4, \bar{2}')  \\ 
 	&- (3 \leftrightarrow 4) + 	\Gamma^\Lambda(\bar{2}, 4, \bar{1}', 3)\Gamma^\Lambda(3, \bar{1}, 4, \bar{2}') + (3 \leftrightarrow 4)
 	\bigg]G^\Lambda(\omega_3)\partial_\Lambda G^\Lambda(\omega_4)\\
 	& \overset{(\text{I})}{=}- \frac{1}{2\pi} \sum_{3, 4} \bigg[
 	\Gamma^\Lambda(3, 4, 1, 2)\Gamma^\Lambda(1', 2', 3, 4)
    - \Gamma^\Lambda(1', 3, 1, 4)\Gamma^\Lambda(4, 2', 3,2) - (3 \leftrightarrow 4)  \\ 
 	&+ 	\Gamma^\Lambda(1', 3, 2, 4)\Gamma^\Lambda(4, 2', 3, 1) + (3 \leftrightarrow 4)
 	\bigg] G^\Lambda(\omega_3)\partial_\Lambda G^\Lambda(\omega_4)\\
 	&\overset{(\text{II})}{=} -\frac{1}{2\pi} \sum_{3, 4} \bigg[
 	\Gamma^\Lambda(3, 4, 1, 2)\Gamma^\Lambda(1', 2', 3, 4)
 	-	\Gamma^\Lambda(1', 4, 1, 3)\Gamma^\Lambda(3, 2', 4,2) - (3 \leftrightarrow 4) \\ 
 	& + 	\Gamma^\Lambda(3, 1',  4, 2)\Gamma^\Lambda(2', 4, 1, 3) + (3 \leftrightarrow 4)
 	\bigg] G^\Lambda(\omega_3)\partial_\Lambda G^\Lambda(\omega_4)
 	= \frac{d}{d\Lambda} \Gamma^\Lambda(1', 2', 1, 2). \label{eq: vertex induction proof}
\end{aligned}
\end{equation}
In step I we applied Eq.~\eqref{eq:vertex-left-to-proof} and transformed the sum indices $\bar{3}$, $\bar{4}$ to 3 and 4 by using that the propagator is odd in frequency space. In step II we exchanged the indices $3 \leftrightarrow 4$ and applied the particle exchange symmetry [Eq.\eqref{eq:particle-exchange}] to the last term. This concludes the proof for the two particle vertex.

Due to the additional vertex components compared to the $\mathfrak{su}(2)$ case, we have to repeat the proof for the self-energy, although we will closely follow Ref.~\cite{buessen2019}. To this end, we first rewrite the relations in Eqs.~(\ref{eq:particle-exchange}, \ref{eq:vertex-H-TR}) for the parametrized vertex, but using natural frequencies
\begin{align*}
	\Gamma^{\Lambda \mu\kappa\lambda}_{i_1 i_2}(\omega_{1'}, \omega_{2'}, \omega_1, \omega_2) &= \Gamma^{\Lambda \mu\lambda\kappa}_{i_2 i_1}(\omega_{2'}, \omega_{1'}, \omega_2, \omega_1) \\
	\Gamma^{\Lambda \mu\kappa\lambda}_{i_1 i_2}(\omega_{1'}, \omega_{2'}, \omega_1, \omega_2) &= \xi(\kappa) \xi(\lambda) \Gamma^{\Lambda \mu\kappa\lambda}_{i_1 i_2}(\omega_1, \omega_2, \omega_{1'}, \omega_{2'}),
\end{align*}
which directly implies
\begin{align}
	\Gamma^{\Lambda \mu\kappa\lambda}_{i_1 i_2}(\omega_1, \omega_2, \omega_1, \omega_2)&=0  \quad \text{if}\quad \xi(\kappa)\xi(\lambda)  = -1 \label{eq:non-local-vertex}\\
	\Gamma^{\Lambda \mu\kappa\lambda}_{i_1 i_1}(\omega_1, \omega_2, \omega_2, \omega_1) &=\xi(\kappa)\xi(\kappa) \Gamma^{\Lambda \mu\lambda\kappa}_{i_1 i_1}(\omega_1, \omega_2, \omega_2, \omega_1). \label{eq:local-vertex}
\end{align}
Using these relations, we can simplify the self-energy flow equation
\begin{equation*}
    \begin{aligned}
         &2\pi \frac{d\Sigma(1', 1)}{d\Lambda}\\
         & \overset{\mathrm{(I)}}{=} \delta_{w_{1'} w_1}\delta_{i_{1'} i_1}  \int d\omega_2 \sum_{\mu,\kappa,\eta} \sum_{s_2, l_2}
	    \Bigg[ \sum_{i_2}\Gamma_{i_1i_2}^{\Lambda\mu\kappa\lambda}(\omega_1, \omega_2, \omega_1, \omega_2)
	    \theta_{s_{1'}s_{1}}^\mu \theta_{s_2s_2}^\mu \theta_{l_{1'}l_{1}}^\kappa \theta_{l_2l_2}^\lambda
	    \\
	    & -\Gamma_{i_1i_1}^{\Lambda\mu\kappa\lambda}(\omega_1, \omega_2, \omega_2, \omega_1)
	    \theta_{s_{1'}s_{2}}^\mu \theta_{s_2s_1}^\mu \theta_{l_{1'}l_{2}}^\kappa \theta_{l_2l_1}^\lambda \Bigg]S^{\Lambda}(\omega_{2}) \\
	    &\overset{\mathrm{(II)}}{=}\delta_{w_{1'} w_1}\delta_{i_{1'} i_1}\delta_{s_{1'}s_{1}}\int d\omega_2\sum_{l_2}
	    \Bigg[ 
	    2\sum_{i_2}\Gamma_{i_1i_2}^{\Lambda0\kappa\lambda}(\omega_1, \omega_2, \omega_1, \omega_2)\theta^\kappa_{l_1'l_1}\theta^\lambda_{l_2l_2} \\
	    &-\sum_{\mu} \Bigg( 
	    \sum_{\kappa>\lambda>0}
	    \Gamma_{i_1i_1}^{\Lambda\mu\kappa\lambda}(\omega_1, \omega_2, \omega_2, \omega_1)
	    (\theta_{l_{1'}l_{2}}^\kappa \theta_{l_2l_1}^\lambda + \theta_{l_{1'}l_{2}}^\lambda \theta_{l_2l_1}^\kappa)
	    + \sum_{\kappa>0} 
	     \Gamma_{i_1i_1}^{\Lambda\mu\kappa0}(\omega_1, \omega_2, \omega_2, \omega_1)
	    (\theta_{l_{1'}l_{2}}^\kappa \theta_{l_2l_1}^0 - \theta_{l_{1'}l_{2}}^0 \theta_{l_2l_1}^\kappa)\\
	    &+ \sum_{\kappa}
	    \Gamma_{i_1i_1}^{\Lambda\mu\kappa\kappa}(\omega_1, \omega_2, \omega_2, \omega_1)
	    \theta_{l_{1'}l_{2}}^\kappa \theta_{l_2l_1}^\kappa 
	    \Bigg)\Bigg] S^{\Lambda}(\omega_{2})\\
	    &\overset{\mathrm{(III)}}{=} \delta_{w_{1'} w_1}\delta_{i_{1'} i_1}\delta_{s_{1'}s_{1}} \delta_{l_{1'}l_1}\int d\omega_2
	    \Bigg[4\sum_{i_2} \Gamma_{i_1i_2}^{\Lambda000}(\omega_1, \omega_2, \omega_1, \omega_2) 
	    - \sum_{\mu,\kappa}\Gamma_{i_1i_1}^{\Lambda\mu\kappa\kappa}(\omega_1, \omega_2, \omega_2, \omega_1)
	    \Bigg]S^{\Lambda}(\omega_{2}).
    \end{aligned}
\end{equation*}
In step I we simply wrote out the self-energy flow equation using the vertex parametrization. In step II we performed the sum over $s_2$ using $(\theta^\mu)^2 = \mathds{1}$ and $\mathrm{Tr}\, \theta^\mu = 2 \delta_{\mu,0}$ and applied Eq.~\eqref{eq:local-vertex} to rearrange the sum over $\kappa$ and $\lambda$. In step III we perform the sum over $l_2$ again by using $\mathrm{Tr}\, \theta^\lambda = 2 \delta_{\lambda,0}$ and  $(\theta^\kappa)^2 = \mathds{1}$ and the (anti)commutation relations $[\theta^\kappa, \theta^0] = 0$ and $\{\theta^\kappa, \theta^\lambda\} = 2 \delta_{\kappa\lambda}$ (for $\kappa, \lambda >0$). In this form of the flow equation it is clear that the self-energy indeed stays diagonal during the flow and, as all two-particle vertex components appearing in the last line are real and the single-scale propagator is imaginary, the self-energy is completely imaginary. That the self-energy is also odd in frequency space can easily be seen by using Eq.~\eqref{eq:full-vertex-conjugate} and the particle exchange symmetry [Eq.\eqref{eq:particle-exchange}]. This concludes the proof for the self-energy.
%%%%%%%%%%%%%%%%%%%%%%%%%%%%%%%%%%%%%%%%%%%%%%%%%%%%%%%%%%
\end{document}